\documentclass[journal]{IEEEtran}
\usepackage{float}
\usepackage{amsmath,amssymb,amsfonts}
\usepackage{graphicx}
\usepackage{textcomp}
\usepackage{colortbl}

\usepackage{adjustbox}
\usepackage{caption}
\usepackage{tabularx}
\usepackage{import}
\usepackage{algpseudocode,algorithm}
\usepackage{listings}
\usepackage{multirow}
\usepackage{caption}
\usepackage{subcaption}
\usepackage{hyperref}
\usepackage{circuitikz}
\usepackage{pgf}
\usepackage{pgfplots}
\usepackage{xcolor}
\usepackage{bm}
\usepackage{pgfplotstable}
\pgfplotsset{compat=1.8}

\usepackage{adjustbox}
\usepackage{changepage}
\usepackage{dblfloatfix}
\usepackage{pythontex}

\usepackage[utf8]{inputenc}
\usepackage{flushend}

\newcommand\xqed[1]{%
  \leavevmode\unskip\penalty9999 \hbox{}\nobreak\hfill
  \quad\hbox{#1}}
\newcommand\exampleEnd{\xqed{$\blacksquare$}}

\ifCLASSOPTIONcompsoc
  \usepackage[nocompress]{cite}
\else
  \usepackage{cite}
\fi

\ifCLASSINFOpdf
\else
\fi
\usepackage{flushend}

\begin{document}
%
\title{Information Leakage through Physical Layer Supply Voltage Coupling Vulnerability}


\author{Sahan~Sanjaya,~Aruna~Jayasena,
        and~Prabhat~Mishra
\\ {\small  University of Florida, Gainesville, Florida, USA}
}

\maketitle

\begin{abstract}
Side-channel attacks exploit variations in non-functional behaviors to expose sensitive information across security boundaries. Existing methods leverage side-channels based on power consumption, electromagnetic radiation, silicon substrate coupling, and channels created by malicious implants. Power-based side-channel attacks are widely known for extracting information from data processed within a device while assuming that an attacker has physical access or the ability to modify the device. In this paper, we introduce a novel side-channel vulnerability that leaks data-dependent power variations through physical layer supply voltage coupling (PSVC). Unlike traditional power side-channel attacks, the proposed vulnerability allows an adversary to mount an attack and extract information without modifying the device. We assess the effectiveness of PSVC vulnerability through three case studies, demonstrating several end-to-end attacks on general-purpose microcontrollers with varying adversary capabilities. These case studies provide evidence for the existence of PSVC vulnerability, its applicability for on-chip as well as on-board side-channel attacks, and how it can eliminate the need for physical access to the target device, making it applicable to any off-the-shelf hardware. Our experiments also reveal that designing devices to operate at the lowest operational voltage significantly reduces the risk of PSVC side-channel vulnerability.
\end{abstract}


%
\IEEEpeerreviewmaketitle

\pagestyle{plain}

\section{Introduction} \label{sec:introduction}

Side-channel leakage represents a significant security concern with the diversity of computing devices used for security-sensitive applications. An adversary can exploit various side-channels, including power consumption, electromagnetic (EM) emanation, and silicon substrate coupling as well as channels created by malicious implants (e.g., hardware Trojans). For example, the covert avenue for information extraction can occur when the power consumption of a device fluctuates during its operation, indirectly revealing valuable insights about the data and cryptographic processes that are being executed. In essence, an adversary can deduce the information processed by a device by closely monitoring and analyzing its power consumption characteristics. 

\begin{figure}[htp]
    \begin{center}
        \small
        \input{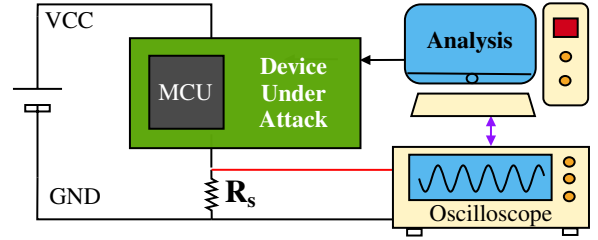}
    \end{center}
      \caption{Overview of a power-based side-channel attack. The current flow through the device is measured using a shunt resistor (\textbf{R\textsubscript{S}}) connected in series to the device, which translates to the power consumption of the device. The power traces provide insights into the ongoing computation in the device.}
    \vspace{-0.15in}
      \label{fig:existing}
\end{figure}

To conduct traditional side-channel attacks, several requirements must be met. First and foremost, the adversary requires physical access to the target device, which may be a microcontroller unit (MCU), an integrated circuit (IC), an embedded system, or even a server. This physical access allows the adversary to monitor the power consumption in real-time with minor modifications to the existing hardware.  However, if current sensors are already available in the device, the adversary can directly probe into the sensors. Figure~\ref{fig:existing} illustrates a typical power side-channel attack setup, where the adversary probes across the current sensor (shunt resistor R\textsubscript{S}). Then by manipulating the inputs to the design, the adversary is able to correlate the computation with the observed power trace of the device. Usually, the initial proof-of-concept (PoC) is constructed offline with a spare device with the same specification as the original device that the adversary is planning to attack. The final attack can be launched based on the PoC on the actual target device and the computations that happened in the device can be deduced in real-time during execution. 

In this paper, we introduce a new side-channel that exposes data-dependent power variations through physical layer supply voltage coupling (PSVC) vulnerability. Unlike traditional power side-channel attacks, our method enables an adversary to extract information without requiring modifications to the device. We evaluate the effectiveness of PSVC vulnerability through three case studies, illustrating various end-to-end attacks on general-purpose microcontrollers with diverse adversary capabilities. 

\subsection{Threat Model}\label{subsec:Threat}

We assume three levels of adversary capabilities as \textit{\textbf{level-1}}, \textit{\textbf{level-2}} and \textit{\textbf{level-3}}, as illustrated by Table~\ref{tab:advcapabilities}.  With the \textit{level-1} capability, the adversary is able to modify the device under attack (e.g. install a shunt resister, modify the firmware, etc.). Attack models relying on side-channel leakage via power consumption typically fall under \textit{level-1} category. However, \textit{level-1} category capabilities may not be practical in most scenarios. Adversaries with \textit{\textbf{level-2}} capabilities cannot make modifications to the hardware but are able to probe into the power supply lines of the victim device. Finally, the most constrained adversary is the \textit{\textbf{level-3}}, where they don't have modification ability (\textit{level-1}) or physical access (\textit{level-2}). Instead in the case of \textit{\textbf{level-3}}, an adversary has access to the wireless communication of the victim device, such as Wi-Fi, Bluetooth, etc. Specifically, we assume that the adversary is situated within the wireless communication range of the victim device.  
\begin{table}[H]
\centering
\caption{Adversary levels based on their ability to access the victim device. To launch a side-channel attack using PSVC, the minimum required adversary capability level is \textbf{\textit{level-3}}.}\label{tab:advcapabilities}
\begin{tabular}{|c|l|}
\hline
\textit{\textbf{Level}} & \textbf{Adversary Capabilities} \\ \hline
\textit{\textbf{1}} & Able to modify the victim device \\ \hline
\textit{\textbf{2}} & Has physical access, cannot modify \\ \hline
\textit{\textbf{3}} & Wireless access within a certain range \\ \hline
\end{tabular}
\end{table}

\subsection{Physical Side-Channel Evaluation}

There are a wide variety of methods for design-time as well as run-time side-channel analysis to reveal secrets. The goal of the design time techniques, such as test vector leakage assessment~\cite{gilbert2011testing, jayasena2023tvla, zhang2021psc, pundir2022power, he2019rtl}, is to detect potential power side-channel leaky designs at the early stages of the design cycle. A vast majority of run-time methods analyze variations in power, current, or path delay to evaluate information leakage. However, these methods are not applicable in many scenarios since they rely on physical access (e.g., to probe power lines) or even minor modifications (e.g., inserting a shunt register) of the device. There are approaches for remote attacks that exploit various side-channels, such as electromagnetic emanation~\cite{agrawal2003side, longo2015soc}, radio frequency emissions~\cite{camurati2018screaming}, and video-based cryptanalysis~\cite{nassi2023video}. However, these methods evaluate computation inside the target MCU itself or data transmissions within chip interconnects. In contrast, our proposed PSVC vulnerability evaluation methodology explores information leakage across voltage domains as well as through wireless carrier signals without modification of the device.

\subsection{Research Contributions}

We introduce a new dimension of physical side-channel vulnerability based on PSVC that can be exploited to launch a wide variety of attacks, including on-chip attacks, on-board attacks, and fully remote attacks. Due to the drastic difference between PSVC from other side-channel sources, it requires the development of new techniques for PSVC side-channel extraction as well as utilization of PSVC for information leakage. Specifically, this paper makes the following contributions.

\begin{itemize}
    \item We introduce a new and effective vulnerability for information leakage, referred as physical layer supply voltage coupling (PSVC).
    
    \item We propose an efficient technique to isolate the PSVC signature of the computation from the actual noise and a methodology for evaluating information leakage through PSVC vulnerability.

    \item We perform end-to-end attacks exploiting PSVC vulnerability on two off-the-shelf victim devices supporting different instruction-set architectures.

    \item We show that PSVC signature propagates between voltage domains, and illustrate an on-board attack on a victim MCU from an IC that shares the same power supply.

    \item We show that PSVC signature can propagate with wireless carrier signals, and illustrate a successful end-to-end attack over Bluetooth transmission.

    \item We establish the fidelity of the PSVC vulnerability with respect to voltage variations in the power supply.

    \item We show that operating an MCU at the lowest operational voltage is a better choice in terms of security since it reduces the risk of PSVC side-channel vulnerability. 

    
\end{itemize}

\textit{To the best of our knowledge, our work is the first effort to discover the PSVC as a side-channel source to extract data that is being processed inside the device under attack on off-the-shelf hardware.}

\subsection{Paper Organization}
The remainder of this paper is organized as follows. Section~\ref{sec:background} provides background on various side-channel sources and surveys related efforts on side-channel attacks. Section~\ref{sec:methodology} demonstrates our methodology for the evaluation of side-channel leakage through physical layer supply voltage coupling. Section~\ref{sec:experiments} performs end-to-end attacks on several configurations of off-the-shelf hardware components and shows the effectiveness of the proposed side-channel attack.
Finally, Section~\ref{sec:conclusion} concludes the paper.

\section{Background and Related Work}\label{sec:background}

In this section, we first discuss different physical side-channel sources used to perform attacks. Next, we survey related efforts on side-channel attacks.

\subsection{Physical Side-Channel Sources}
There are different types of side-channels that can leak information about the internal operation of a device. Specifically, we start the discussion with the most popular power side-channels~\cite{nassi2023video} followed by recently introduced silicon substrate coupling~\cite{camurati2018screaming,schellenberg2018remote}. Finally, we introduce supply voltage coupling, which we will be using in this paper as the side-channel leakage source.

\vspace{0.05in}
\noindent \textbf{Side-channels due to power consumption:} The power consumption varies in response to the changing logic states and data processing during computation in a device. These fluctuations in power create distinctive patterns that can be analyzed to infer information about the operations being executed.  Figure~\ref{fig:existing} shows a typical setup for performing power side-channel attacks. Device power consumption is assessed by employing a shunt resistor (denoted as R\textsubscript{S}), while simultaneously adjusting the device's inputs. This process helps establish the relationship between computational activity and power usage. Once the correlation is figured out, an adversary can observe power variations when a similar device is deployed and the adversary is able to deduce sensitive data, such as cryptographic keys or plaintext, without directly accessing the internal memory or processes of the target device. 

\vspace{0.05in}
\noindent \textbf{Side-channels due to silicon substrate coupling:} In the context of electronic systems, silicon substrate coupling refers to the phenomenon where energy or signals traverse between different components and traces within the same silicon substrate, whether by magnetic fields inducing voltage, electric fields enabling energy transfer, or through direct connections. This phenomenon can lead to undesirable electromagnetic interference and radio-frequency interference, which an adversary can use to recover sensitive information about the operations being computed inside the device~\cite{camurati2018screaming,schellenberg2018remote}. Designers try to make use of the effects of coupling to optimize the performance and reliability of devices.

\vspace{0.05in}
\noindent \textbf{Side-channels due to supply voltage coupling:}  
In this paper, we introduce the tolerable noise that is coupled with the power signature of the dominant component of the device as \textbf{physical layer supply voltage coupling} (PSVC) and utilize it as the physical side-channel source to leak information from the power dominant component. 
In order to launch an attack by exploiting PSVC vulnerability, the minimum adversary capabilities must be at \textit{\textbf{level-3}}. PSVC adversary capabilities are similar to the attack methods that rely on side-channel leakage due to silicon substrate coupling. It is worth noting that silicon substrate coupling-based methods have the ability to extract data that is only in transit within chip interconnects or data processed within the wireless transmitter itself. 
\textit{To the best of our knowledge, the concept of using physical layer supply voltage coupling as a side-channel on off-the-shelf hardware has not been explored in the existing literature.}

\subsection{Related Work}

A vast majority of research on side-channel attacks relies on power side-channel analysis that measures a device's power consumption by monitoring variations in its current consumption during internal computations. Since the focus of this paper is on side-channel vulnerabilities that extend beyond the direct measurement of current consumption, in this section, we survey methods that exploit radio frequency emissions as well as visual feedback to uncover potential security risks and avenues for information leakage.

A side-channel attack based on the standard power LED's brightness and the color variations is proposed in~\cite{nassi2023video}. In practice, power LEDs are typically connected to the embedded device, serving to offer a visual indication to the user that the device is powered and operational. The underlying concept is that minor power variations introduced during the computation will be reflected in the LED and can be captured visually. Therefore, the authors have used a video-based cryptanalysis to recover the secret information that was processed inside the embedded system.








The side-channel attack based on silicon substrate coupling is illustrated in~\cite{camurati2018screaming}. The authors illustrate the effect of silicon substrate coupling leaking the information computed in mixed-signal chips.  While this specific attack is constrained to chips housed on the same die, the authors demonstrate their ability to extract confidential data from the digital processor by analyzing the frequency variations of an analog radio frequency device that coexists on the same silicon substrate.

Danieli et al.~\cite{danieli2023revealing} propose a simple power analysis technique to retrieve data from peripheral communication such as serial communication (UART), JTAG, communication to memory, and flash drives. It can only extract information traveling through interconnects and unable to retrieve any information from the processing cores. The author's primary focus is on extracting information from on-board signals, both with and without a galvanic connection to the TX antenna chip. In other words, the attack is capable of retrieving data traveling through the same die chip interconnects and unable to retrieve information that is being processed in the processing core.



\begin{figure*}[htp]
    \begin{center}
    \vspace{-0.1in}
        \input{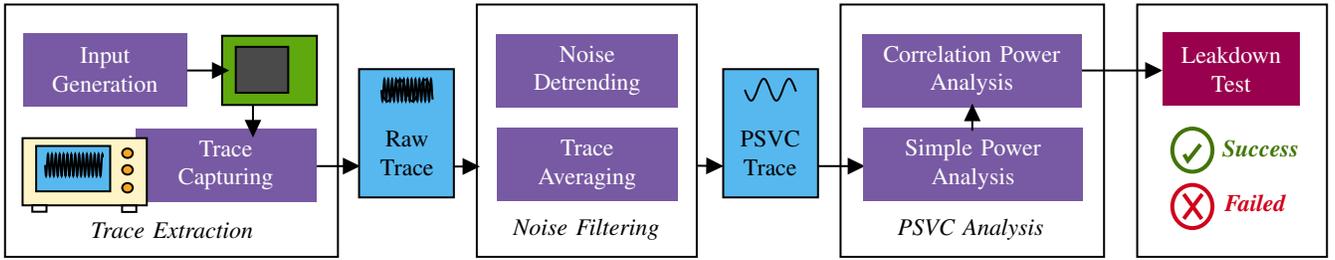}
    \end{center}
      \caption{Overview of our side-channel analysis framework utilizing physical layer supply voltage coupling (PSVC). This framework consists of four major tasks: trace extraction from the victim device, noise filtering to extract PSVC traces, PSVC trace analysis, and leakdown test to determine whether the device under attack is revealing any secret.}
    \vspace{-0.1in}
      \label{fig:overview}
\end{figure*}

Field-Programmable Gate Array (FPGA)-based side-channel attacks that rely on a hardware Trojan introduced during the manufacturing process are proposed in~\cite{schellenberg2021inside,schellenberg2018remote}. FPGA Power Distribution Network (PDN) is a complex and unique network of power supply connections and pathways within an FPGA, responsible for delivering regulated power to gate arrays while ensuring signal integrity and minimizing electromagnetic interference. During the manufacturing process of the PDN, a malicious hardware Trojan needs to be inserted to perform the side-channel attack and retrieve the information that was being computed inside the FPGA.

The scope of current methods is constrained by three significant drawbacks. First, they often necessitate physical or visual proximity to the target device under attack. Second, these approaches frequently demand the introduction of malicious hardware modifications, such as shunt resistors~\cite{flynn2018one} or hardware Trojans~\cite{jordan2018big,hudson2019modchips,shwartz2020inner}. Lastly, remote methods that do not rely on hardware alterations struggle to recover data that is actively being computed within the other components of the device under attack. In contrast, our approach leverages off-the-shelf devices that share the same power supply and their existing architectures to extract confidential information without necessitating any hardware modifications.


\section{Evaluation of PSVC Vulnerability for Information Leakage}\label{sec:methodology}

In this section, we present our methodology to evaluate hardware devices against side-channel leakage due to physical layer supply voltage coupling (PSVC). Figure~\ref{fig:overview} provides an overview of the proposed methodology that consists of four major tasks. The first task generates inputs, feeds them into the device under attack, and captures raw traces. The second task filters the noise to isolate the PSVC signature from traces obtained by running the test vectors on the device under attack. The third task performs simple power analysis as well as correlation power analysis of the collected PSVC traces. The fourth task performs leakdown test to determine the success rate of the attack. We first outline the problem formulation. Next, we describe each of the four tasks in detail.

\subsection{Problem Formulation}\label{subsec:problem}

During the design phase, every integrated circuit (IC) is engineered with specific tolerances for both input voltage and noise variations, ensuring that the IC's functionality remains unaffected within these defined ranges. The manufacturer will specify these values in their application notes (datasheet) with recommended capacitor configurations and layout design at the IC supply voltage points. This information assists printed circuit board (PCB) designers in effectively incorporating these values into their designs. The practical reason behind the coupling between the supply voltage noise and the data being processed can be explained as follows. Power is hierarchically distributed and undergoes voltage regulation at multiple stages, supplying various chips on the board. Inside the chip, a mesh-like network powers individual transistors, with integrated resistance ($R$), capacitance ($C$), and inductance ($L$) components, either by design or as parasitic elements\cite{arabi2007power}.

\begin{figure*}[htp]
    \begin{center}
    \vspace{-0.1in}
        \small
        \input{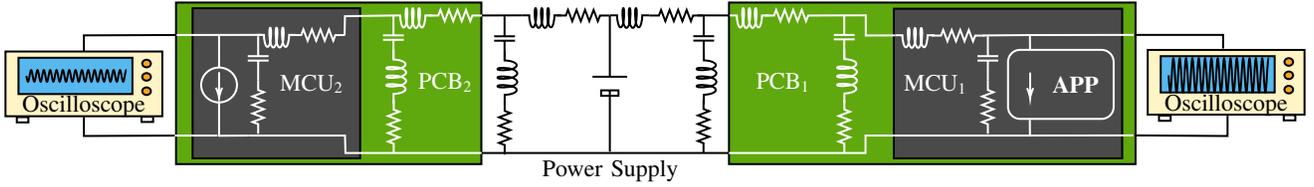}
    \end{center}
      \vspace{-0.15in}
      \caption{Power distribution system model at different hierarchical levels of hardware components. On-chip current demand propagates through the power rails into the board-level power rails. Here the power signature of MCU\textsubscript{1} when running an application program (APP) will be coupled with the power rails of the entire system. This will affect other components that share the same power rails and MCU\textsubscript{2} will see a power signature of MCU\textsubscript{1} running the APP.} 
    \vspace{-0.1in}
      \label{fig:powerDistribution}
\end{figure*}

Figure~\ref{fig:powerDistribution} illustrates a simplified diagram of a device with two MCU chips sharing the same power supply. Here MCU\textsubscript{1} is running a more power-intensive application (APP) and MCU\textsubscript{2} is running a general computation. During MCU operation, transistors switch based on the processed data, resulting in varying power consumption ($P=V\times$I), which leads to fluctuations in current. These current changes, described by the equation $\Delta V = L \times \frac{dI}{dt}+ IR$, cause voltage fluctuations on the power rails due to inductive and resistive elements within the device. These voltage fluctuations are supposed to be mitigated by the smoothing capacitors recommended by the manufacturer~\cite{kocher1999differential}. However, when designers use these manufacturer-specified decoupling capacitor configurations, still a tolerable amount of noise will escape. In the case of the example system in Figure~\ref{fig:powerDistribution}, the power signature of APP running on MCU\textsubscript{1} will be present in the power rails as noise. This tolerable noise does not affect the operation of MCU\textsubscript{2} and other components present in the system. As this noise is still coupled with the power signature of the APP running on MCU\textsubscript{1}, MCU\textsubscript{2} will be able to monitor the power signature of MCU\textsubscript{1} and able to deduce what is happening inside the APP. Information leakage that happens due to the above phenomena can be categorized under the scope of PSVC vulnerability. Figure~\ref{fig:overview} illustrates the four main steps involved in evaluating PSVC vulnerability and the following four sections describe each step in detail. 

\subsection{Trace Extraction from Victim Device}\label{subsec:traceextraction}

The first step for trace extraction is to start with a victim device that we want to evaluate against PSVC side-channel leakage. Before obtaining the PSVC trace from the selected device, we need to generate known input values to be fed into the application program (APP in Figure~\ref{fig:powerDistribution}) that we intend to recover secrets. For ease of illustration, let us assume that the APP is an Advanced Encryption Standard (AES) cryptographic algorithm from the AESLib~\cite{davy2015aes} library with AES-128-bit electronic codebook (ECB) implementation. We first describe the generation of input patterns. Next, we discuss the collection of PSVC traces.

\vspace{0.1in}
\noindent \textbf{Randomized Input Generation: }
Our goal is to formulate inputs in a manner that amplifies the correlation between the PSVC trace and the input. Ensuring an even distribution across the keyspace is crucial in this process because if a correlation exists between the key and the PSVC trace, uniformly manipulating the keyspace can accentuate the trace data, thereby highlighting the correlation. In the case of AES, we fix the key and randomize the plaintext that is used for encryption. Once the input data is written into the internal memory of the device, we proceed to the next step.

\vspace{0.1in}
\noindent \textbf{Trace Collection: } 
In this step, the objective is to capture a PSVC trace from the victim's device. Note that the PSVC trace is embedded into the noise on the power rails. Therefore, PSVC carrying noisy trace can be collected in two ways:

\begin{itemize}
    \item Use an oscilloscope to probe directly into victim device power rails (\textit{VCC} and \textit{GND}).
    \item Using Software Defined Radio (SDR) to capture the carrier signal of the device under attack.
\end{itemize}

The first method requires physical access to the device or its power supply (e.g., wall charger or battery) to probe directly into the power lines. The next method is applicable if the device under attack consists of a radio transmitter (e.g., Bluetooth or WiFi). If automated trace alignment is required, a trigger pin can be assigned from the hardware device to trigger the trace-capturing device. However, this step is optional since there are signal processing tools that can perform the required alignment on collected trace data.

\subsection{Noise Filtering and PSVC Extraction}\label{subsec:noisefiltering}

The trace extracted from the device is dominated by the noise introduced by different components in the circuit. We need to isolate the PSVC trace from the noise to evaluate the effect of information leakage. For this purpose, we use two noise filtering techniques: noise detrending and trace averaging.

\begin{figure}[htp]
\begin{subfigure}{0.5\linewidth}
\begin{center}
\includegraphics[width=\textwidth]
{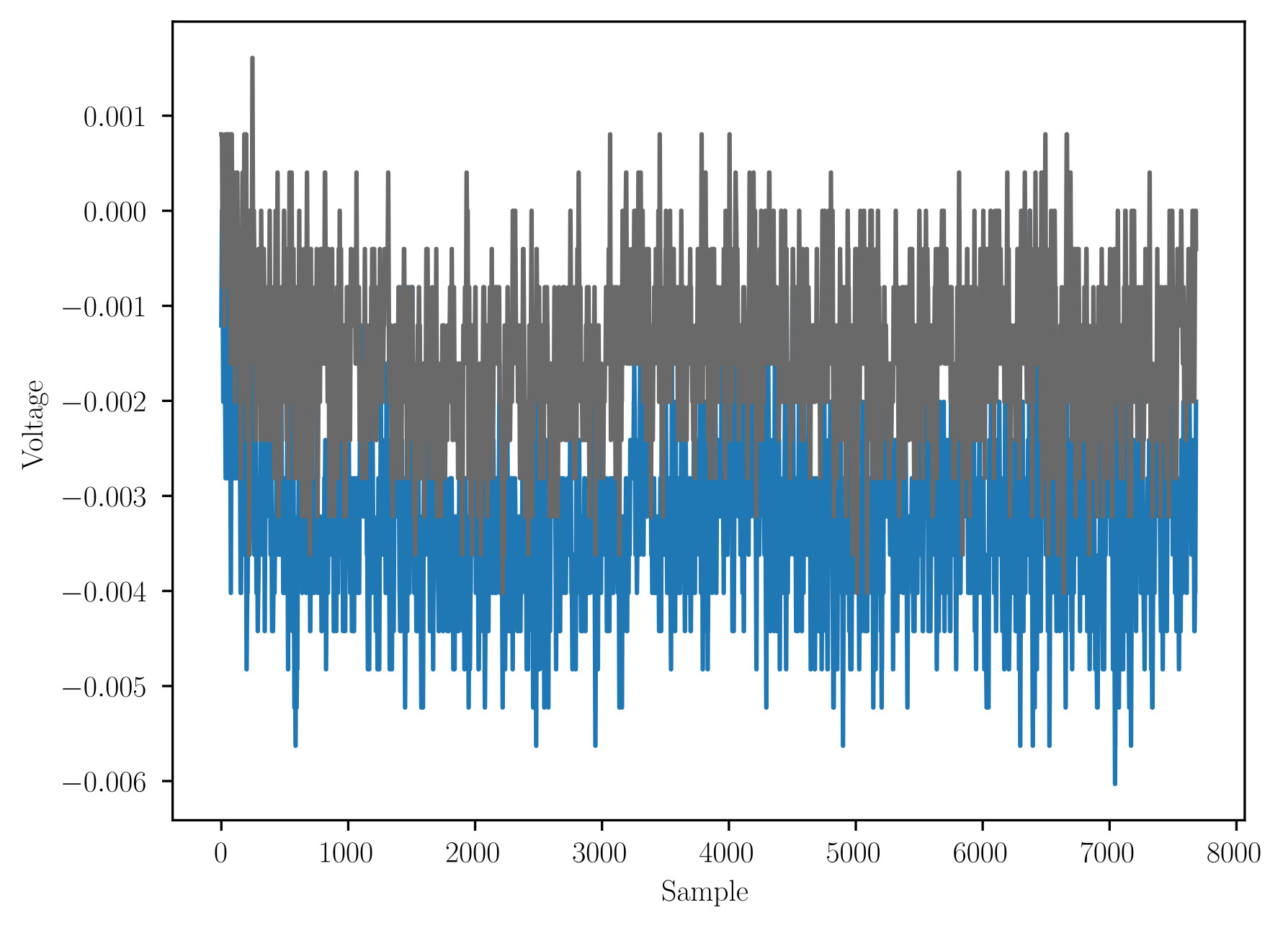}
\end{center}
\vspace{-0.1in}
\caption{Before detrending operation}\label{subfig:before}
\end{subfigure}%
\begin{subfigure}{0.5\linewidth}
\begin{center}
\includegraphics[width=\textwidth]
{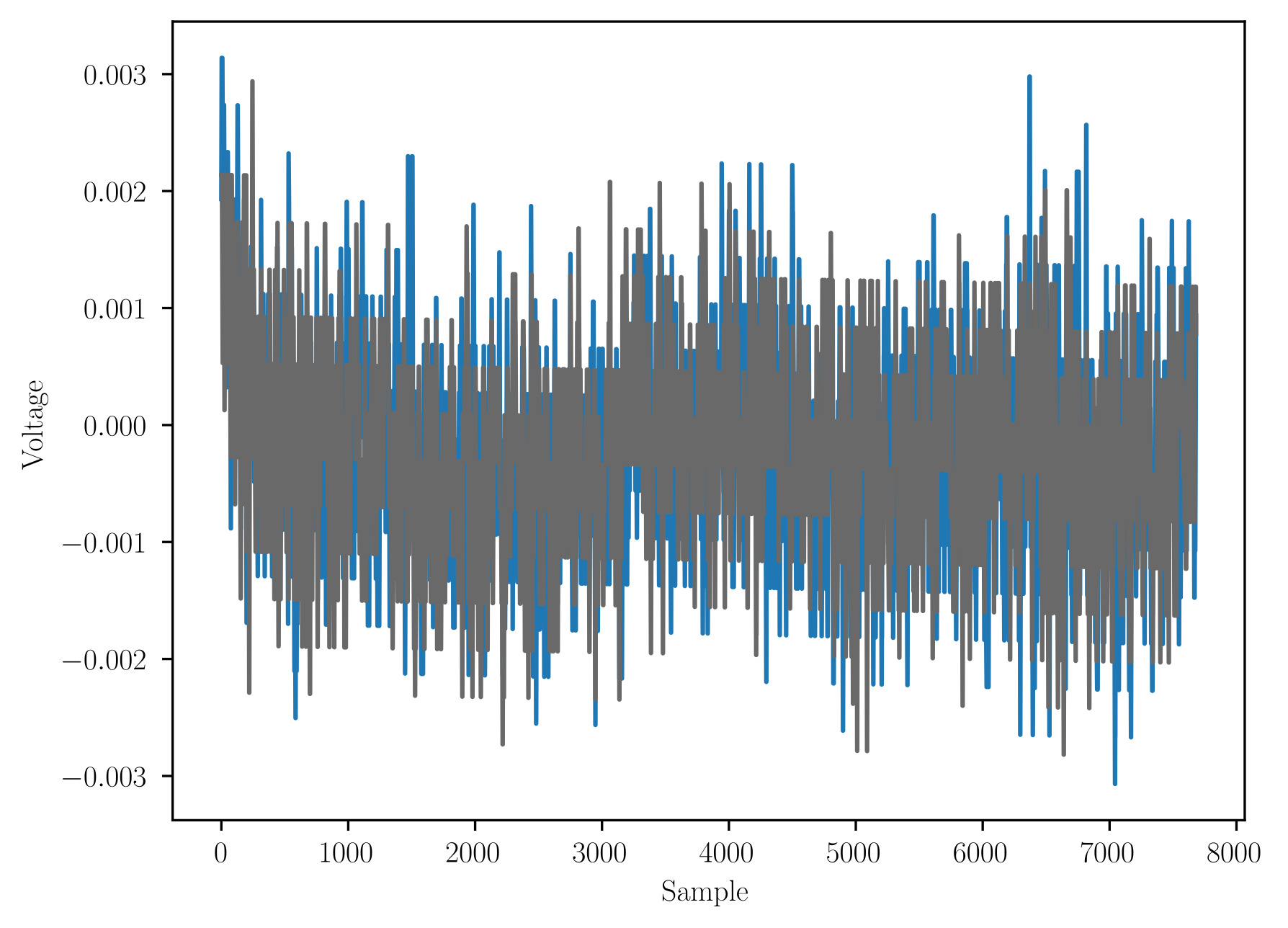}
\end{center}
\vspace{-0.1in}
\caption{After detrending operation}\label{subfig:after}
\end{subfigure}
\hspace{-0.2in}
\caption{Noise removal with detrending operation on traces collected over AES block cipher.}
\vspace{-0.1in}
\label{fig:detrend}
\end{figure}

\vspace{0.1in}
\noindent \textbf{Noise Detrending: } The term detrending refers to the process of removing a trend from a signal. In a noisy environment (with a noisy power supply), random unwanted trends often appear in the captured traces. There could be multiple sources for these unwanted trends such as the switching noise generated by a switching power supply, environmental noise, and noise from the measuring device itself. These unwanted trends affect each trace differently, resulting in a misalignment of the amplitude of the signal (y-axis). To isolate the PSVC trace, we follow a two-step process. First, we perform linear detrending by modeling and subtracting linear trends present in the signal, and then the signal becomes more evident. Finally, we use high-pass or low-pass filtering to remove noise within designated frequency bands, preserving the inherent signal frequencies of the hidden PSVC signature. Figure~\ref{fig:detrend} illustrates an instance of a collected trace from the AES ECB mode, before (Figure~\ref{subfig:before}) and after (Figure~\ref{subfig:after}) performing the noise detrending. It can be observed that the variations in the trace due to noise are removed after performing the detrending operation.

\vspace{0.1in}
\noindent \textbf{Trace Averaging: } In the process of highlighting the PSVC trace from the collected trace data, we need to consider the Signal to Noise Ratio (SNR). Here the signal is the effect of PSVC. Our objective is to improve the signal (effect of PSVC) while reducing the effect of noise. In other words, the higher the SNR ratio, the better we isolate the PSVC signature from the noise. To improve the SNR ratio, we perform trace averaging. The number of traces considered for the averaging depends on the specific scenario and configurations of the device under attack. Therefore, we define the $Avg(N)$ function in Equation~\ref{eqn:avgfunc} to denote the trace averaging function where $\bar{X}(t)$ is the average trace at time $t$, $N$ is the total number of traces,
$X_n(t)$ is the value of the $n^{th}$ trace at time $t$. The summation runs through all $N$ traces, and $t$ represents a specific time point.
\begin{gather}\footnotesize
\bar{X}(t) = \frac{1}{N} \sum_{n=1}^{N} X_n(t) \nonumber \\
Avg(N) = \{\bar{X}(0), \bar{X}(1), \ldots ,\bar{X}(T)\}
\label{eqn:avgfunc}
\end{gather}

\begin{figure}[htp]
\begin{subfigure}{0.5\linewidth}
\begin{center}
\includegraphics[width=\textwidth]
{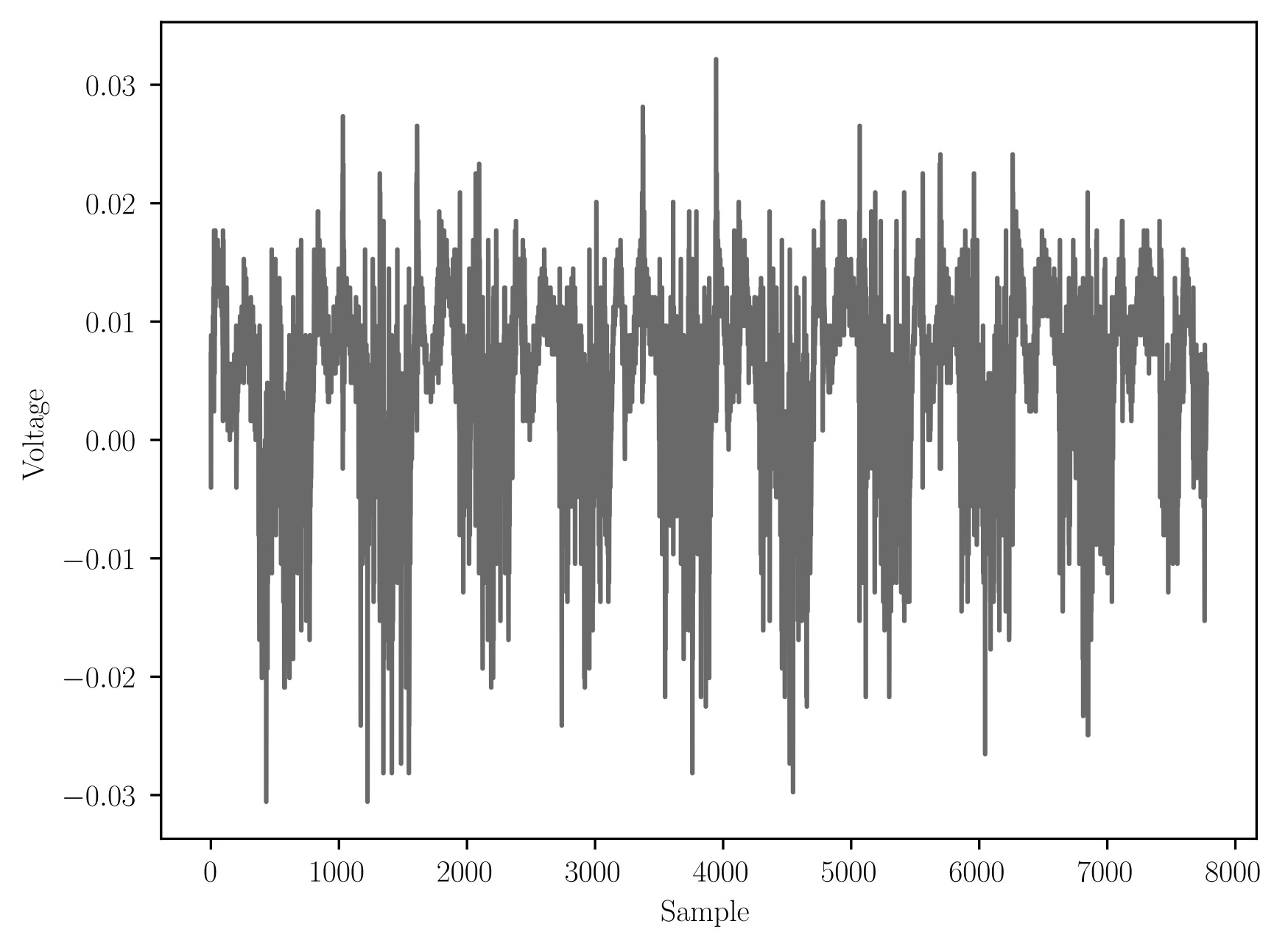}
\end{center}
\vspace{-0.1in}
\caption{Before averaging ($Avg(1)$)}\label{subfig:abefore}
\end{subfigure}%
\begin{subfigure}{0.5\linewidth}
\begin{center}
\includegraphics[width=\textwidth]
{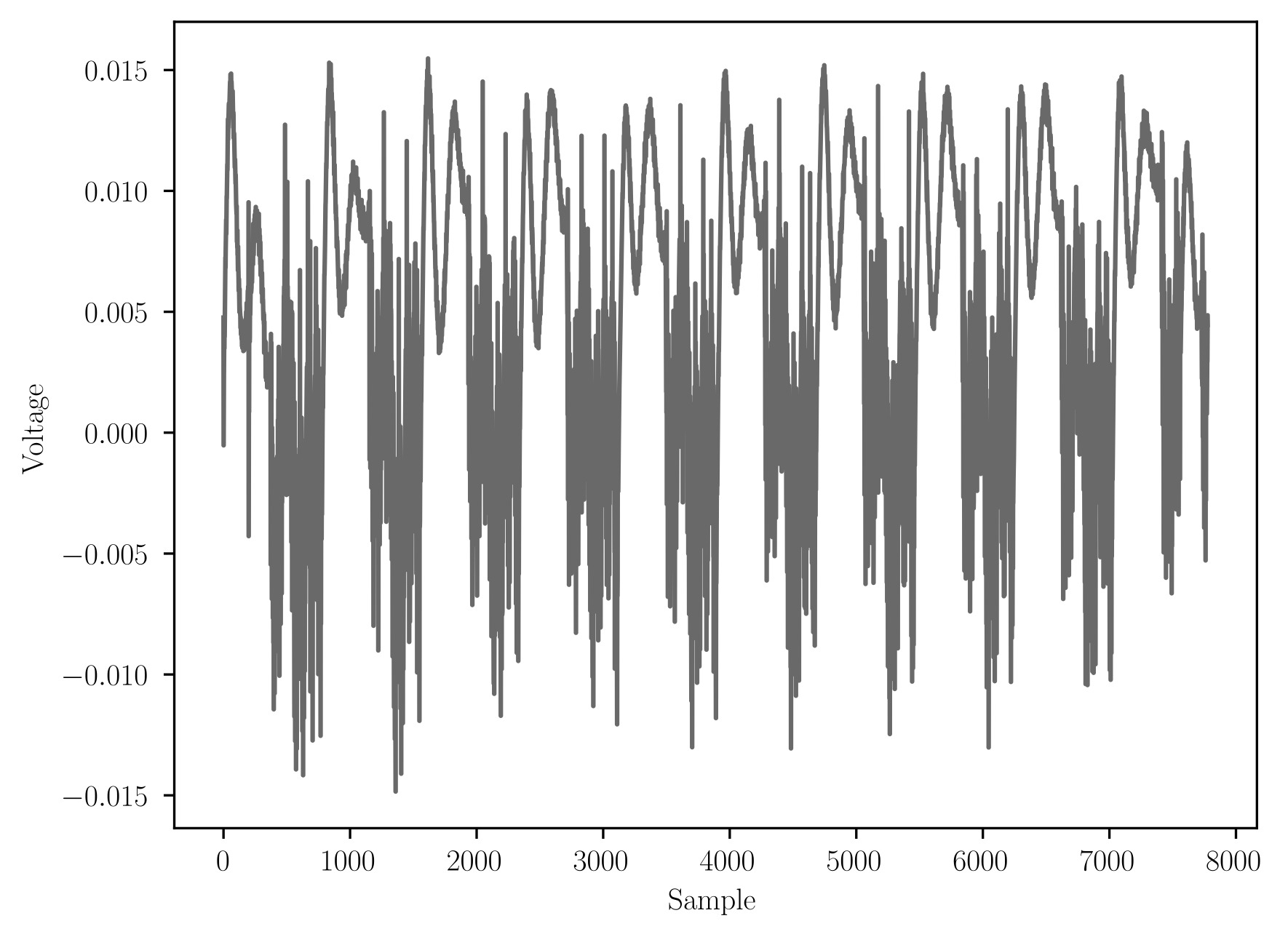}
\end{center}
\vspace{-0.1in}
\caption{After averaging ($Avg(10)$)}\label{subfig:aafter}
\end{subfigure}
\hspace{-0.2in}
\caption{Improving the SNR with trace averaging on traces collected over AES block cipher on \textit{BlackPill}. Trace averaging is able to improve SNR by eliminating random DC shifts caused by the power supply.}
\vspace{-0.2in}
\label{fig:average}
\end{figure}

Table~\ref{tab:SNR_vs_average} presents the variation of the SNR with respect to different numbers of averaged traces of $Avg(N)$ of AES block cipher running on an \textit{Arduino nano} victim device. When the number of traces ($N$) increases, the SNR improves significantly. Figure~\ref{fig:average} presents an instance of trace average on the AES block cipher. It can be observed that compared to one trace ($Avg(1)$) in Figure~\ref{subfig:abefore}, an average of $N=10$ traces ($Avg(10))$ is superior with respect to SNR in Figure~\ref{subfig:aafter} with eliminated random DC shifts caused by the power supply.

\begin{table}[htp]
\centering
\begin{tabular}{|l|c|c|c|c|}
\hline
$Avg(N)$ & 1 & 2 & 5 & 10 \\ \hline
SNR (dB) & \cellcolor[HTML]{FFE599}5.06 & \cellcolor[HTML]{FFF2CC}8.61 & \cellcolor[HTML]{D9EAD3}14.24 & \cellcolor[HTML]{6AA84F}21.93 \\ \hline
\end{tabular}
    \caption{Improvement of SNR of the PSVC signal with the increase in number of traces ($N$) used for trace averaging, $Avg(N)$. }
    \label{tab:SNR_vs_average}
\end{table}

After executing noise detrending and subsequently implementing trace averaging, we successfully isolated the trace signal pertinent to the PSVC signature of the target device. The next step involves analyzing this refined signal to assess whether the captured PSVC trace has the potential to leak confidential information from the application (APP in Figure~\ref{fig:powerDistribution}) running on the victim device or not.

\subsection{PSVC Trace Analysis}\label{subsec:psvcanalysis}

At this stage, we have an isolated signal trace that encodes the PSVC signature of the device. Next, we need an effective way to process this signal to find a correlation between the input secret values given to the tasks and the PSVC signature. In order to perform this, we explore two techniques: simple power analysis and correlation power analysis.

\vspace{0.1in}
\noindent \textbf{Simple Power Analysis (SPA): } The intention of SPA is that if the device leaks information from the side-channel environment, it might be directly visible on the signature traces. This is a very simple technique that is performed by doing visual inspections on the side-channel traces. During the inspection of PSVC traces, we pay attention to the algorithm of the application program (APP) running on the victim device. If the implementation has obvious drawbacks such as imbalance branch statements or specific hardware components that created specific PSVC signatures during certain operations, their effects can be visually observed on the PSVC trace itself. Note that simple power analysis may not work for all the devices as we illustrate in Section~\ref{sec:experiments}.

\vspace{0.1in}
\noindent \textbf{Correlation Power Analysis (CPA): } When the observation based on SPA does not yield any meaningful results, we can move towards a more experimental method of correlation power analysis. CPA aims to establish a correlation between measured side-channel signature and expected side-channel signature~\cite{brier2004correlation}. CPA has the ability to reveal correlations even if noise is not completely filtered from the trace. CPA takes two inputs, expected value and observed value, and produces the output of correlation value between the observed and expected value. The expected value can be calculated in the following two ways from the randomized input values generated in Section~\ref{subsec:traceextraction}.

\begin{itemize}
    \item \textit{Hamming Weight Model}: PSVC signature is correlated with the number of $1$'s in the selected data, that is being computed in the device.
    \item \textit{Hamming Distance Model}: Consecutive PSVC signatures are correlated with the number of transitions from $1 \rightarrow 0$ and $0 \rightarrow 1$  in each bit of the two consecutive selected data.
\end{itemize}

We utilize the most simple and straightforward Hamming weight model for the correlation power analysis. Then we compute the Pearson's correlation coefficient $r_{xy}$ between the expected value $x$ and observed value $y$ from Equation~\ref{eqn:pearsoneq}. Here, $\bar{x}$ and $\bar{y}$ are average of samples $x$ and $y$,  respectively. 
\begin{gather}\label{eqn:pearsoneq}
    r_{xy} = \frac{\sum_{i=1}^{n} (x_i - \bar{x})(y_i - \bar{y})}{\sqrt{\sum_{i=1}^{n} (x_i - \bar{x})^2 \sum_{i=1}^{n} (y_i - \bar{y})^2}}
\end{gather}

We extract an array of values from each trace at time $t$, representing the observed values $y$. The expected value array $x$ comprises Hamming weight values for each plaintext corresponding to a guessed key byte value. Subsequently, we compute $r_{xy}$ for every time point and each key byte value and store the absolute maximum correlation value for each key byte value in an array denoted as $r$. The array $r$ reflects the correlation between the guessed secret key values assumed to be used in the application (APP) of the victim device and the PSVC power trace. The next step involves classifying these correlation values to determine whether the device under test is indeed leaking information through the PSVC side-channel.

 \vspace{0.1in}
\noindent \textbf{Example 1 (CPA)}: \textit{Suppose $p_1 = [0.01, 0.2, 0.05, 0.1]$, $p_2 = [0.02, 0.1, 0.03, 0.2]$  and $p_3 = [0.01, 0.2, 0.04, 0.2]$  are three traces observed after trace extraction and noise filtering stages. Assume that $p_1$, $p_2$, and $p_3$ are captured for a fixed secret key and three different plain texts \texttt{0x02,0x05} and \texttt{0x01}, respectively. Let us guess the secret key as \texttt{0x03} and choose selected data as the xor output between plain text and the guessed key to calculate the Hamming weight. Then, for $t = 0$,  we get $x = [0.01, 0.02, 0.01]$, $y = [1, 6, 2]$, and $r_{xy} = 0.98$. For $t = 1$, $x = [0.2,0.1,0.2]$, $y = [1, 6,2]$, and $r_{xy} = -0.98$. Likewise, after calculating $r_{xy}$ for all time sampling points for three different key guesses (\texttt{0x01,0x02} and \texttt{0x03}), we get $r = [0.69, 0.99, 0.98]$.} \exampleEnd

\subsection{Leakdown Test}

It is important to clarify that a higher absolute correlation value does not guarantee the accuracy of the key guess; it simply makes it more likely. Therefore, we implement the final step as the leakdown test where we perform key guess validation with correlation thresholding and conclude whether the attack is a ``success'' or ``failed''. We define a distance value as shown in Equation~\ref{eqn:distfunc}, denoted as $d$, which measures the difference between the highest correlation value and the average correlation values in $r$. 
\begin{gather}\label{eqn:distfunc}
    d = max(r) - \frac{1}{256} \sum_{j=1}^{256} r[j] 
    \vspace{-0.3in}
\end{gather}

If the $d$ value for a guess key value is greater than a predefined threshold $\lambda$, we can say that the guessed key byte value is a correct key guess. This signifies that the attack is successful. However, it is important to strike a balance with the threshold, as setting it too low can result in false positives (accepting incorrect key guesses) while setting it too high can lead to false negatives (rejecting correct key guesses). The value of the threshold is based on empirical experiments and involves some trial and error. The threshold value is directly affected by the characteristics of the device under attack and the quality of the trace-capturing device. 

 \vspace{0.05in}
\noindent \textbf{Example 2 (Leakdown Test)}: \textit{Distance values for guessed key in Example 1 are as follows: for key guess \texttt{0x01}, $d = -0.20$; for key guess \texttt{0x02}, $d = 0.10$; and for key guess \texttt{0x03}, $d = 0.09$. If we define $\lambda = 0.095$ based on historical data, we can conclude that \texttt{0x02} is the secret key.} \exampleEnd
 \vspace{-0.05in}


    
\section{Experiments} \label{sec:experiments}

To demonstrate the applicability and effectiveness of information leakage through physical layer voltage coupling (PSVC), we have performed four case studies utilizing two hardware boards and two trace collection methods. 

\begin{figure}[h]
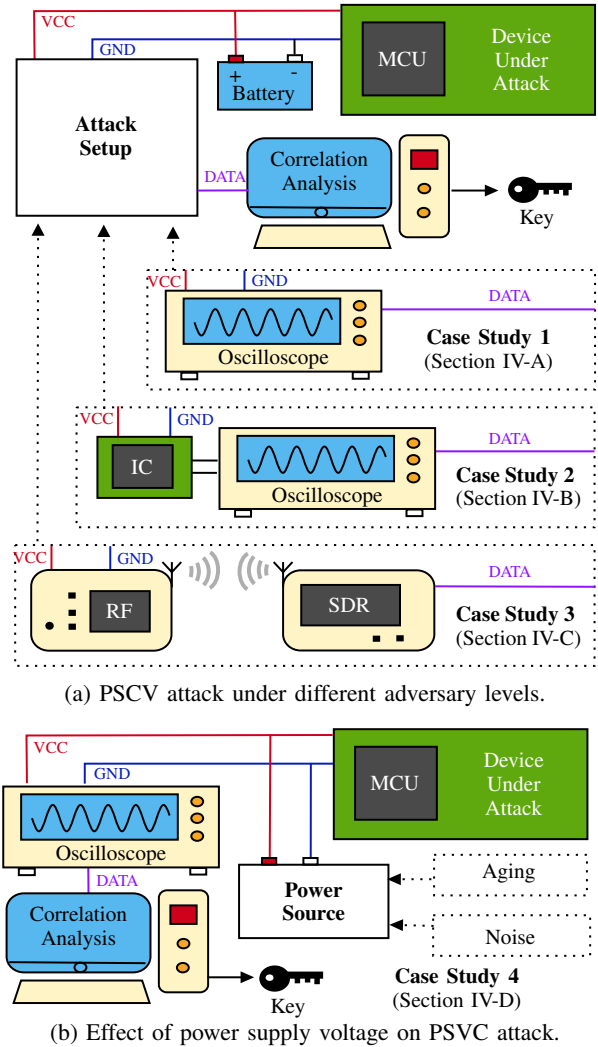

  
    \begin{subfigure}{1\linewidth}
      \centering
        \input{fig/evaluation}
      \caption{PSCV attack under different adversary levels.}
      \label{fig:cases1-3}
      \vspace{0.1in}
    \end{subfigure}%

    \begin{subfigure}{1\linewidth}
      \centering
  \input{fig/batteryeval}
      \vspace{-0.1in}
      \caption{Effect of power supply voltage on PSVC attack.}
      \label{fig:battery}

    \end{subfigure}%
     \caption{Experimental setup for four case studies. Case Study 1 assumes direct access to the victim device, while Case Study 2 indirectly (via another IC) connects to the victim device.  Case Study 3 is the ultimate end-to-end attack that can be launched using PSVC, where the adversary is in most constrained \textbf{\textit{level-3}}. Case Study 4 explores the effect of power supply voltage on PSVC vulnerability.}
    \vspace{-0.15in}
      \label{fig:evaluation}
\end{figure}

\vspace{0.1in}
\noindent \textbf{Four Case Studies:} The first three case studies explore the possibility of launching a PSVC side-channel attack on three different configurations. We have selected three different configurations of devices with increasing complexity as three case study scenarios (Section~\ref{subsec:case_1} to Section~\ref{subsec:case_3}). Figure~\ref{fig:cases1-3} shows the overview of the device configurations used in the first three case studies.  Case Study 1 directly connects the oscilloscope inputs to the power lines (VCC and GND) and the oscilloscope output (DATA) to the computer for PSVC trace (correlation) analysis. In contrast, Case Study~2 connects the oscilloscope to an IC, which is connected to the device under attack. In other words, Case Study 2 represents the scenario in Figure~\ref{fig:powerDistribution}. Note that Case Study 3 is the ultimate attack where the adversary is in \textit{\textbf{level-3}} and only needs wireless proximity to the device.
All these three case studies will have three things in common: the power supply, victim MCU, and the PSVC evaluation framework as discussed in Section~\ref{sec:methodology}.
The fourth case study evaluates the effect of the PSVC side-channel attack under different power supply configurations in Section~\ref{subsec:case_4}. The different power supply configurations used for Case Study 4 are illustrated in Figure~\ref{fig:battery}.

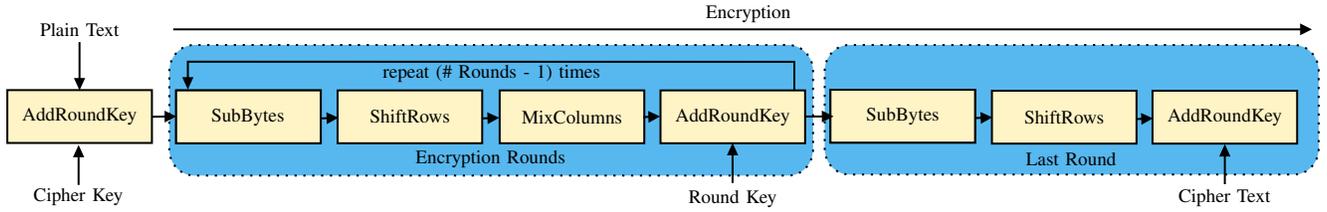
\begin{figure*}[htp]
    \begin{center}
    \vspace{-0.1in}
        \tikzset{every picture/.style={line width=0.75pt}} 

\begin{tikzpicture}[x=0.75pt,y=0.75pt,yscale=-1,xscale=0.95]

\draw  [fill={rgb, 255:red, 87; green, 184; blue, 239 }  ,fill opacity=1 ][dash pattern={on 0.84pt off 2.51pt}] (435.99,34.13) .. controls (435.99,26.91) and (441.85,21.05) .. (449.07,21.05) -- (685.17,21.05) .. controls (692.39,21.05) and (698.25,26.91) .. (698.25,34.13) -- (698.25,73.36) .. controls (698.25,80.59) and (692.39,86.44) .. (685.17,86.44) -- (449.07,86.44) .. controls (441.85,86.44) and (435.99,80.59) .. (435.99,73.36) -- cycle ;
\draw  [fill={rgb, 255:red, 87; green, 184; blue, 239 }  ,fill opacity=1 ][dash pattern={on 0.84pt off 2.51pt}] (88.07,34.24) .. controls (88.07,26.96) and (93.97,21.05) .. (101.25,21.05) -- (416.64,21.05) .. controls (423.93,21.05) and (429.83,26.96) .. (429.83,34.24) -- (429.83,73.78) .. controls (429.83,81.06) and (423.93,86.97) .. (416.64,86.97) -- (101.25,86.97) .. controls (93.97,86.97) and (88.07,81.06) .. (88.07,73.78) -- cycle ;
\draw  [fill={rgb, 255:red, 255; green, 244; blue, 198 }  ,fill opacity=1 ] (2.06,43.83) -- (78.83,43.83) -- (78.83,70.51) -- (2.06,70.51) -- cycle ;

\draw  [fill={rgb, 255:red, 255; green, 244; blue, 198 }  ,fill opacity=1 ] (91.68,43.98) -- (168.46,43.98) -- (168.46,70.66) -- (91.68,70.66) -- cycle ;
\draw  [fill={rgb, 255:red, 255; green, 244; blue, 198 }  ,fill opacity=1 ] (91.68,43.98) -- (168.46,43.98) -- (168.46,70.66) -- (91.68,70.66) -- cycle ;

\draw  [fill={rgb, 255:red, 255; green, 244; blue, 198 }  ,fill opacity=1 ] (263.45,43.98) -- (340.22,43.98) -- (340.22,70.66) -- (263.45,70.66) -- cycle ;

\draw  [fill={rgb, 255:red, 255; green, 244; blue, 198 }  ,fill opacity=1 ] (177.4,43.98) -- (254.17,43.98) -- (254.17,70.66) -- (177.4,70.66) -- cycle ;

\draw  [fill={rgb, 255:red, 255; green, 244; blue, 198 }  ,fill opacity=1 ] (348.83,43.98) -- (425.6,43.98) -- (425.6,70.66) -- (348.83,70.66) -- cycle ;

\draw  [fill={rgb, 255:red, 255; green, 244; blue, 198 }  ,fill opacity=1 ] (439.05,43.66) -- (515.83,43.66) -- (515.83,70.34) -- (439.05,70.34) -- cycle ;
\draw  [fill={rgb, 255:red, 255; green, 244; blue, 198 }  ,fill opacity=1 ] (439.05,43.66) -- (515.83,43.66) -- (515.83,70.34) -- (439.05,70.34) -- cycle ;

\draw  [fill={rgb, 255:red, 255; green, 244; blue, 198 }  ,fill opacity=1 ] (610.03,44.16) -- (686.81,44.16) -- (686.81,70.84) -- (610.03,70.84) -- cycle ;

\draw  [fill={rgb, 255:red, 255; green, 244; blue, 198 }  ,fill opacity=1 ] (524.77,44.16) -- (601.54,44.16) -- (601.54,70.84) -- (524.77,70.84) -- cycle ;

\draw    (78.76,57.28) -- (88.78,57.15) ;
\draw [shift={(91.78,57.11)}, rotate = 179.27] [fill={rgb, 255:red, 0; green, 0; blue, 0 }  ][line width=0.08]  [draw opacity=0] (6.25,-3) -- (0,0) -- (6.25,3) -- cycle    ;
\draw    (40.4,19.6) -- (40.4,40.98) ;
\draw [shift={(40.4,43.98)}, rotate = 270] [fill={rgb, 255:red, 0; green, 0; blue, 0 }  ][line width=0.08]  [draw opacity=0] (6.25,-3) -- (0,0) -- (6.25,3) -- cycle    ;
\draw    (97.9,40.54) -- (97.9,29.49) -- (419.88,29.54) -- (419.88,43.82) ;
\draw [shift={(97.9,43.54)}, rotate = 270] [fill={rgb, 255:red, 0; green, 0; blue, 0 }  ][line width=0.08]  [draw opacity=0] (6.25,-3) -- (0,0) -- (6.25,3) -- cycle    ;
\draw    (39.9,91.72) -- (39.91,74.5) ;
\draw [shift={(39.91,71.5)}, rotate = 90.04] [fill={rgb, 255:red, 0; green, 0; blue, 0 }  ][line width=0.08]  [draw opacity=0] (6.25,-3) -- (0,0) -- (6.25,3) -- cycle    ;
\draw    (425.71,57.17) -- (437.73,57.12) ;
\draw [shift={(440.73,57.11)}, rotate = 179.78] [fill={rgb, 255:red, 0; green, 0; blue, 0 }  ][line width=0.08]  [draw opacity=0] (6.25,-3) -- (0,0) -- (6.25,3) -- cycle    ;
\draw    (90.13,13.17) -- (691.64,13.17) ;
\draw [shift={(694.64,13.17)}, rotate = 180] [fill={rgb, 255:red, 0; green, 0; blue, 0 }  ][line width=0.08]  [draw opacity=0] (6.25,-3) -- (0,0) -- (6.25,3) -- cycle    ;
\draw    (168.92,58.11) -- (174.8,57.89) ;
\draw [shift={(177.8,57.78)}, rotate = 177.85] [fill={rgb, 255:red, 0; green, 0; blue, 0 }  ][line width=0.08]  [draw opacity=0] (6.25,-3) -- (0,0) -- (6.25,3) -- cycle    ;
\draw    (254.34,58.11) -- (260.21,57.89) ;
\draw [shift={(263.21,57.78)}, rotate = 177.85] [fill={rgb, 255:red, 0; green, 0; blue, 0 }  ][line width=0.08]  [draw opacity=0] (6.25,-3) -- (0,0) -- (6.25,3) -- cycle    ;
\draw    (339.75,57.44) -- (345.63,57.22) ;
\draw [shift={(348.62,57.11)}, rotate = 177.85] [fill={rgb, 255:red, 0; green, 0; blue, 0 }  ][line width=0.08]  [draw opacity=0] (6.25,-3) -- (0,0) -- (6.25,3) -- cycle    ;
\draw    (516.14,57.94) -- (522.02,57.72) ;
\draw [shift={(525.02,57.61)}, rotate = 177.85] [fill={rgb, 255:red, 0; green, 0; blue, 0 }  ][line width=0.08]  [draw opacity=0] (6.25,-3) -- (0,0) -- (6.25,3) -- cycle    ;
\draw    (601.41,58.44) -- (607.28,58.22) ;
\draw [shift={(610.28,58.11)}, rotate = 177.85] [fill={rgb, 255:red, 0; green, 0; blue, 0 }  ][line width=0.08]  [draw opacity=0] (6.25,-3) -- (0,0) -- (6.25,3) -- cycle    ;
\draw    (387.09,90.72) -- (387.1,73.5) ;
\draw [shift={(387.1,70.5)}, rotate = 90.04] [fill={rgb, 255:red, 0; green, 0; blue, 0 }  ][line width=0.08]  [draw opacity=0] (6.25,-3) -- (0,0) -- (6.25,3) -- cycle    ;
\draw    (648.42,91.32) -- (648.43,74.1) ;
\draw [shift={(648.43,71.1)}, rotate = 90.04] [fill={rgb, 255:red, 0; green, 0; blue, 0 }  ][line width=0.08]  [draw opacity=0] (6.25,-3) -- (0,0) -- (6.25,3) -- cycle    ;

\draw (258.6,78.35) node  [xscale=0.9,yscale=0.9] [align=left] {\begin{minipage}[lt]{231.66pt}\setlength\topsep{0pt}
\begin{center}
{\footnotesize Encryption Rounds}
\end{center}

\end{minipage}};
\draw (566.86,78.63) node  [xscale=0.9,yscale=0.9] [align=left] {\begin{minipage}[lt]{178.17pt}\setlength\topsep{0pt}
\begin{center}
{\footnotesize Last Round}
\end{center}

\end{minipage}};
\draw (648.28,98.17) node  [xscale=0.9,yscale=0.9] [align=left] {\begin{minipage}[lt]{53.06pt}\setlength\topsep{0pt}
\begin{center}
{\footnotesize Cipher Text}
\end{center}

\end{minipage}};
\draw (40.45,13.17) node  [xscale=0.9,yscale=0.9] [align=left] {\begin{minipage}[lt]{53.06pt}\setlength\topsep{0pt}
\begin{center}
{\footnotesize Plain Text}
\end{center}

\end{minipage}};
\draw (259.15,35.54) node  [xscale=0.9,yscale=0.9] [align=left] {\begin{minipage}[lt]{218.59pt}\setlength\topsep{0pt}
\begin{center}
{\footnotesize repeat (\# Rounds - 1) times}
\end{center}

\end{minipage}};
\draw (39.81,97.84) node  [xscale=0.9,yscale=0.9] [align=left] {\begin{minipage}[lt]{41.15pt}\setlength\topsep{0pt}
\begin{center}
{\footnotesize Cipher Key}
\end{center}

\end{minipage}};
\draw (387.21,98.84) node  [xscale=0.9,yscale=0.9] [align=left] {\begin{minipage}[lt]{40.19pt}\setlength\topsep{0pt}
\begin{center}
{\footnotesize Round Key}
\end{center}

\end{minipage}};
\draw (395.47,5.44) node  [xscale=0.9,yscale=0.9] [align=left] {\begin{minipage}[lt]{404.42pt}\setlength\topsep{0pt}
\begin{center}
{\footnotesize Encryption}
\end{center}

\end{minipage}};
\draw (129.74,57.46) node  [xscale=0.9,yscale=0.9] [align=left] {\begin{minipage}[lt]{52.65pt}\setlength\topsep{0pt}
\begin{center}
{\footnotesize SubBytes}
\end{center}

\end{minipage}};
\draw (215.79,56.93) node  [xscale=0.9,yscale=0.9] [align=left] {\begin{minipage}[lt]{53.06pt}\setlength\topsep{0pt}
\begin{center}
{\footnotesize ShiftRows}
\end{center}

\end{minipage}};
\draw (301.84,57.26) node  [xscale=0.9,yscale=0.9] [align=left] {\begin{minipage}[lt]{53.06pt}\setlength\topsep{0pt}
\begin{center}
{\footnotesize MixColumns}
\end{center}

\end{minipage}};
\draw (387.21,57.26) node  [xscale=0.9,yscale=0.9] [align=left] {\begin{minipage}[lt]{53.06pt}\setlength\topsep{0pt}
\begin{center}
{\footnotesize AddRoundKey}
\end{center}

\end{minipage}};
\draw (40.45,56.95) node  [xscale=0.9,yscale=0.9] [align=left] {\begin{minipage}[lt]{53.06pt}\setlength\topsep{0pt}
\begin{center}
{\footnotesize AddRoundKey}
\end{center}

\end{minipage}};
\draw (477.44,56.86) node  [xscale=0.9,yscale=0.9] [align=left] {\begin{minipage}[lt]{53.06pt}\setlength\topsep{0pt}
\begin{center}
{\footnotesize SubBytes}
\end{center}

\end{minipage}};
\draw (563.16,57.36) node  [xscale=0.9,yscale=0.9] [align=left] {\begin{minipage}[lt]{53.06pt}\setlength\topsep{0pt}
\begin{center}
{\footnotesize ShiftRows}
\end{center}

\end{minipage}};
\draw (648.42,57.36) node  [xscale=0.9,yscale=0.9] [align=left] {\begin{minipage}[lt]{53.06pt}\setlength\topsep{0pt}
\begin{center}
{\footnotesize AddRoundKey}
\end{center}

\end{minipage}};

\end{tikzpicture}
    \end{center}
      \caption{Overview of AES encryption algorithm that consists of a series of well-defined steps, including substitution, permutation, and mixing operations, which are known as the SubBytes, ShiftRows, and MixColumns transformations. The AES encryption process consists of multiple rounds. If a device is leaking information via PSVC vulnerability, these individual rounds and their internal steps become visible on the PSVC trace.} 
    \vspace{-0.1in}
      \label{fig:aes}
\end{figure*}

\vspace{0.1in}
\noindent \textbf{Two Hardware Boards:} To demonstrate the applicability of our framework, we have used two types of devices with different instruction-set architectures in each of the configurations: (i) \textit{Arduino nano} development board with \textit{Atmel ATmega328} microprocessor based on \textit{RISC AVR} architecture, and (ii) \textit{BlackPill} development board with \textit{STM32F401} microprocessor based on \textit{ARM} architecture. These two hardware boards are shown in Figure~\ref{fig:method_1}.

\vspace{0.1in}
\noindent \textbf{Two Trace Collection Methods:} For the experiments, PSVC traces were collected using the following two devices. From both trace capturing devices, the data were collected from \textit{Matlab R2020a} API. To process the traces collected in each case study, we have developed scripts in \textit{Python} and \textit{Matlab}. All the algorithms for correlation power analysis were written in Python. Experiments were conducted in a system with 16Gb RAM on an AMD Ryzen~7 processor.

\begin{itemize}
    \item \textit{Keysight  DSOX1102G}~\cite{Keysight} oscilloscope controlled through the Virtual Instrument Software Architecture (VISA) protocol. This trace collection scheme is used for Case Study 1, Case Study 2, and Case Study~4.

    \item HackRF~\cite{HackRF} with CubicSDR~\cite{CubicSDR} software-defined radio (SDR) module using radio frequency (RF) signal capturing ability. This trace collection mechanism is used for Case Study 3.
\end{itemize}

\vspace{0.1in}
\noindent \textbf{Application Program (APP):} We have selected the Advanced Encryption Standard (AES) as the application (APP) running on the MCU for our case studies. We use AES since it is designed to operate on fixed-size data blocks (typically 128 bits), and its internal operations consist of a series of well-defined steps, including substitution, permutation, and mixing operations, which are known as the SubBytes, ShiftRows, and MixColumns transformations as illustrated in Figure~\ref{fig:aes}. AES employs a key schedule to generate round keys for each encryption round followed by an AddRoundKey operation. The encryption process consists of multiple rounds (typically 10, 12, or 14 rounds, depending on the key size 128, 192, and 256 bits, respectively), wherein the data undergoes transformations via these operations. If a device is leaking information via PSVC vulnerability, these individual rounds and their internal steps become visible on the PSVC trace as we demonstrate through four case studies.



\subsection{Case Study 1: Exploit PSVC Vulnerability to Mount an Attack with Power Rail Probing}\label{subsec:case_1}

In this case study, we perform an end-to-end attack exploiting the PSVC vulnerability on the victim device when probe access is available to launch the attack. We set the adversary capability as \textit{\textbf{level-2}} as defined in the threat model (Section~\ref{subsec:Threat}). Here the adversary has physical access to the victim's device but is not allowed to do any modifications to the device. First, we explain the experimental setup used for this case study. Then we perform an end-to-end attack on \textit{Arduino nano} and \textit{BlackPill} development boards and demonstrate the attack capabilities of PSVC-based side-channel attacks with power rail probing.

\subsubsection{Experimental Setup}

In this case study, we utilize SPA followed by CPA to analyze AES encryption. Typically, side-channel attacks focus on modeling power patterns from the first round of computation of AES. The first round provides valuable insights into how plaintext and key bytes are combined.
We focus on the SubBytes operation, chosen for its susceptibility to side-channel attacks. The SubBytes's vulnerability stems from its data-dependent and key-dependent transformations. During this operation, each data byte undergoes substitution with another byte based on a fixed lookup table. As these substitutions occur, the power consumption discloses information about the processed bytes and the encryption key. The inherent data and key dependencies within the SubBytes operation result in observable patterns and correlations in the side-channel data. We leverage this effect to illustrate the PSVC side-channel leakage in this case study.

\begin{figure}[htp]
\begin{subfigure}{0.5\linewidth}
\begin{center}
\vspace{-0.1in}
\input{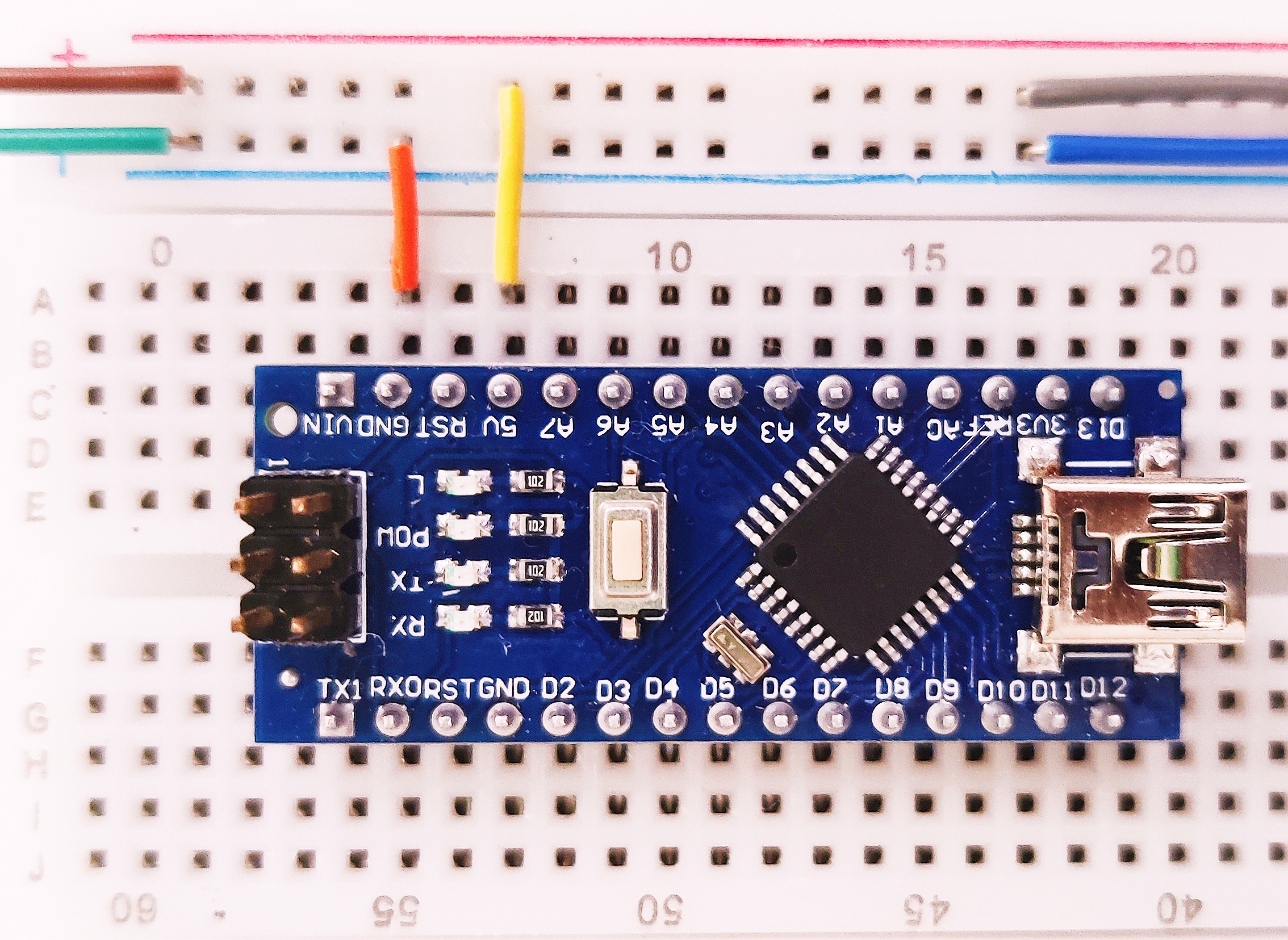}
\end{center}
\vspace{-0.1in}
\caption{RISC based \textit{Arduino nano}}\label{fig:arduinonano}
\end{subfigure}%
\begin{subfigure}{0.5\linewidth}
\begin{center}
\input{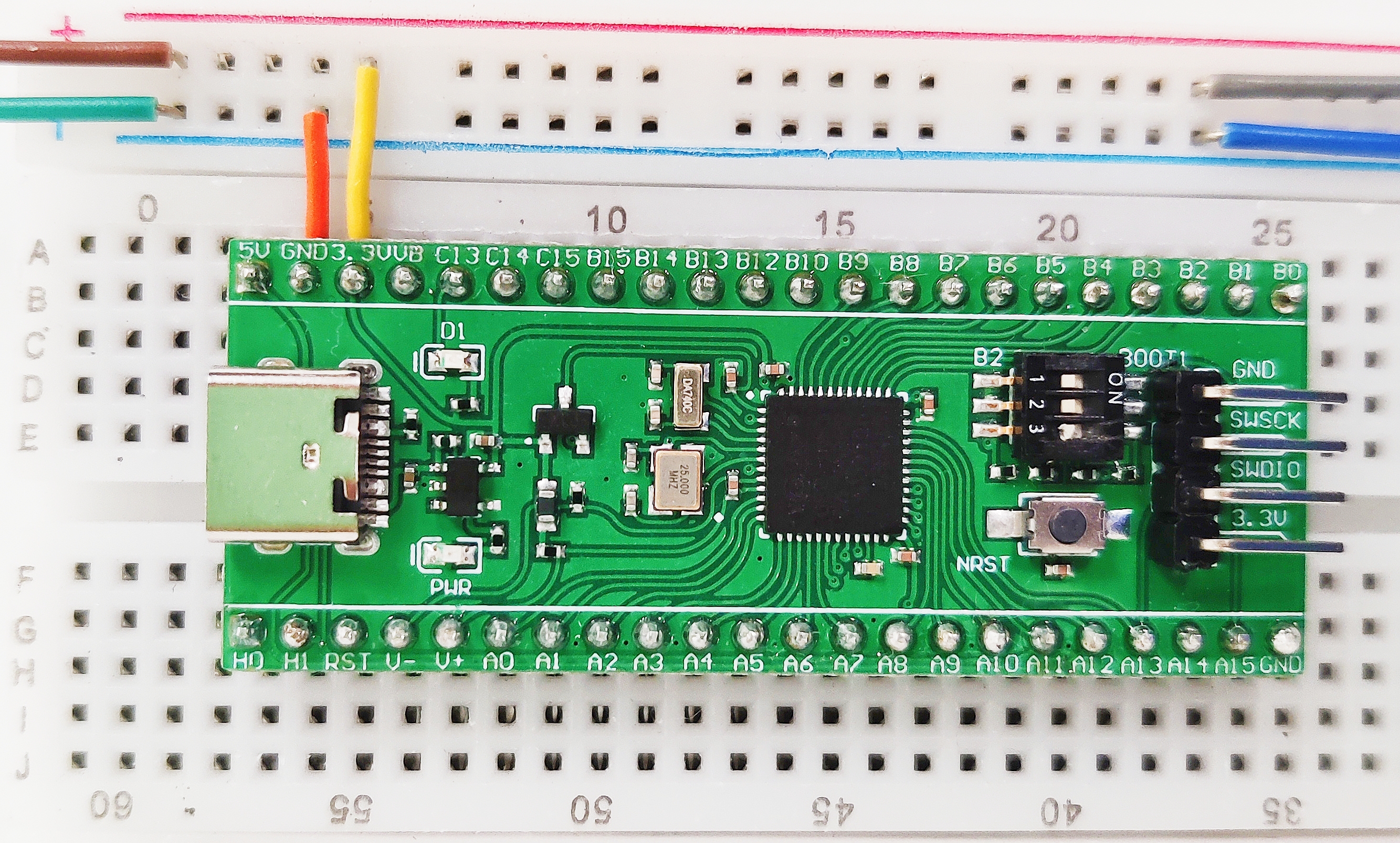}
\end{center}
\vspace{-0.1in}
\caption{ARM based \textit{BackPill}}\label{fig:blackpill}
\end{subfigure}
\caption{Two attack setups used in Case Study 1 with two different development boards with two different instruction set architectures of RISC and ARM.}
\label{fig:method_1}
\end{figure}

\begin{figure*}[htp]
\centering
\include{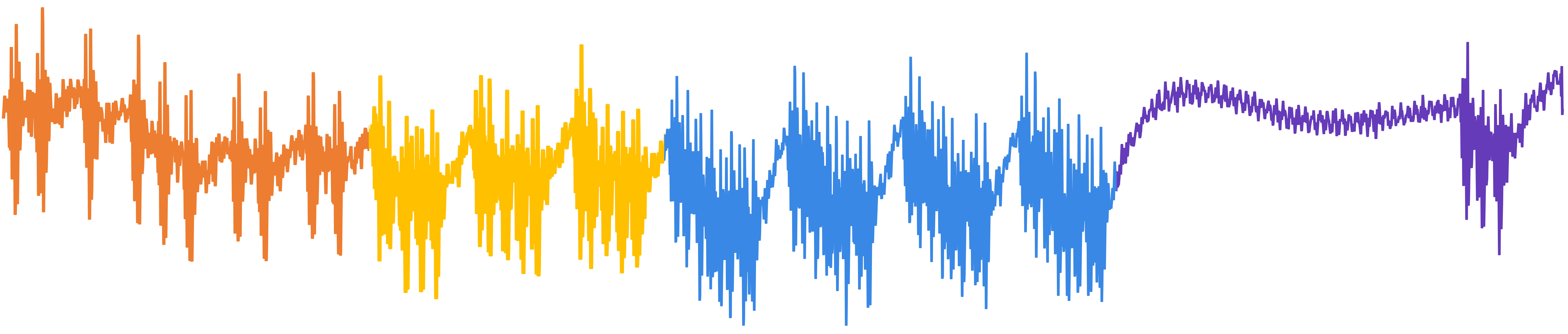}
\vspace{-0.25in}
\caption{Simple power analysis of a PSVC trace for one round of AES-128 captured from \textit{STM32F401} MCU in \textit{BlackPill} development board. Here all operations illustrated within the encryption round of Figure~\ref{fig:aes} can be clearly observed.}
\label{fig:SPA_stm32_1_round}
\vspace{-0.15in}
\end{figure*}

Figure~\ref{fig:method_1} illustrates the two hardware setups used in this case study, these devices are connected to a noisy power supply, which serves as the worst-case power source (lower SNR ratio) for the device under attack. Inside the MCU of the device under attack, we execute an AES encryption repeatedly. In this case study, we have performed two sets of experiments, using two different MCUs: the \textit{Arduino nano} development board with an \textit{Atmel ATmega328} MCU (Figure~\ref{fig:arduinonano}) and the \textit{BlackPill} development board with an \textit{STM32F401} MCU (Figure~\ref{fig:blackpill}). With \textit{\textbf{level-2}} adversary capabilities, we probe the power inputs to the MCU (development board). Subsequently, we capture sufficient traces using an oscilloscope to perform CPA. After the detrend operation on captured traces based on the steps discussed in Section~\ref{subsec:noisefiltering}, we take an average of 10 traces for the \textit{Arduino nano} ($Avg(10)$) and 10 traces for the \textit{BlackPill} ($Avg(10)$) to find better traces. Then, CPA is applied to these filtered traces.

\begin{figure}[htp]
\begin{subfigure}{0.5\linewidth}
\begin{center}
\includegraphics[width=\textwidth]
{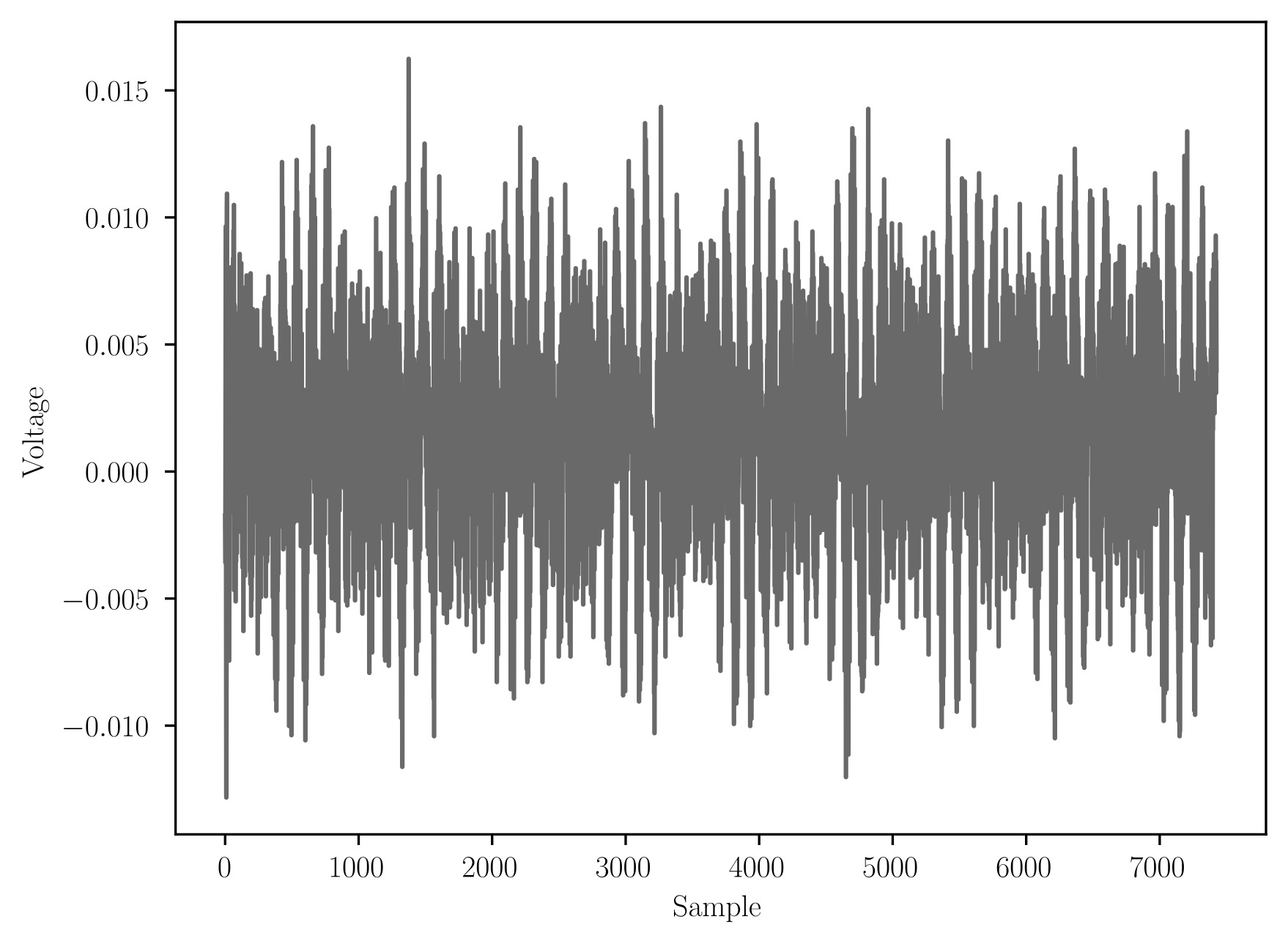}
\vspace{-0.2in}
\end{center}
\vspace{-0.1in}
\caption{PSVC trace on ATmega328}\label{fig:SPA_ard}
\end{subfigure}%
\begin{subfigure}{0.5\linewidth}
\begin{center}
\includegraphics[width=\textwidth]
{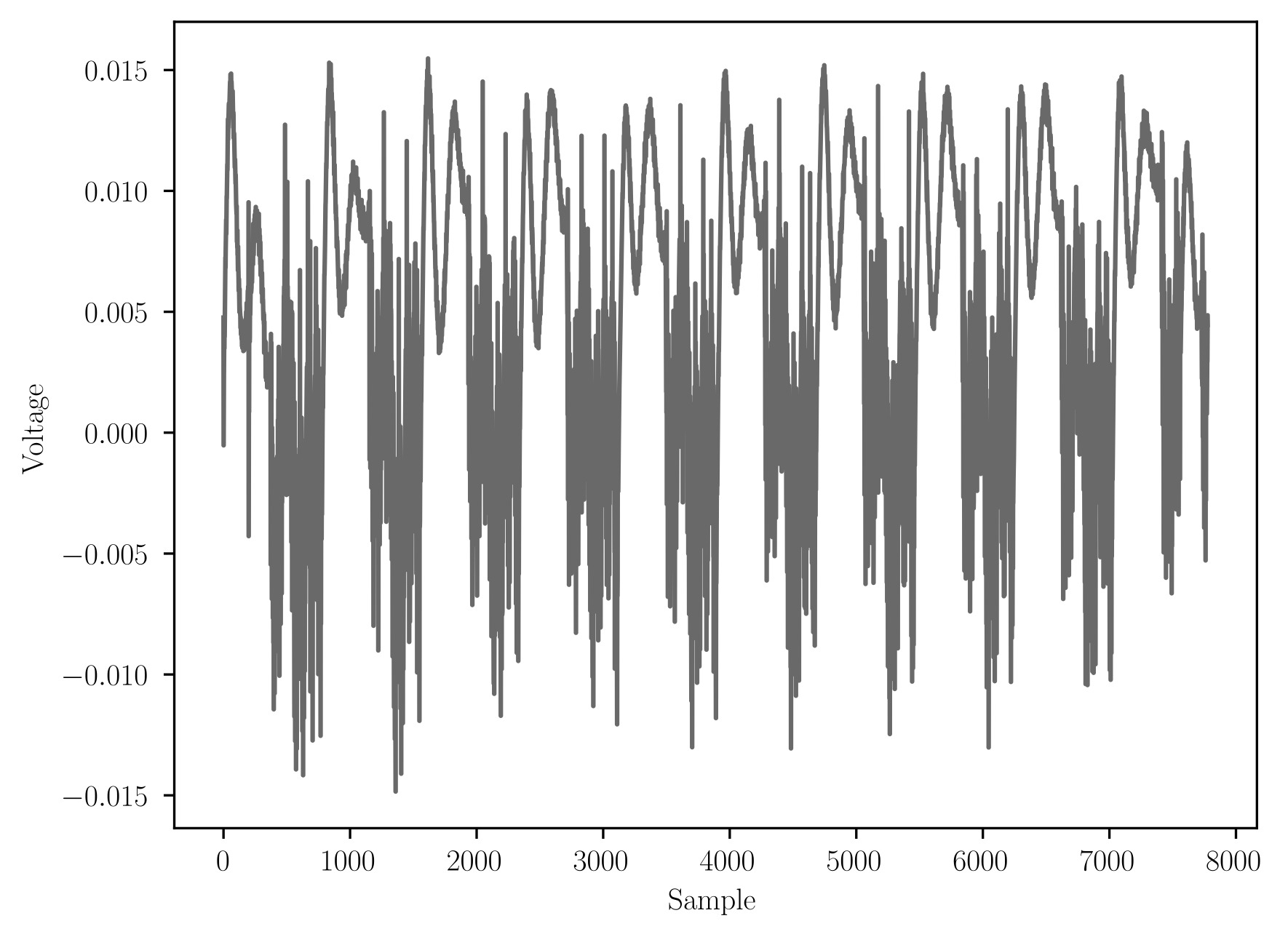}
\vspace{-0.2in}
\end{center}
\vspace{-0.1in}
\caption{PSVC trace on STM32F401}\label{fig:SPA_stm32}
\end{subfigure}
  \vspace{-0.2in}
     \caption{Simple power analysis of PSVC traces for AES-128 on two different MCUs demonstrates the PSVC vulnerability. The traces include the signature of ten AES rounds.}
      \label{fig:case_study_1_SPA}
\end{figure}

\begin{figure*}[h]
\vspace{-0.2in}
\centering
\input{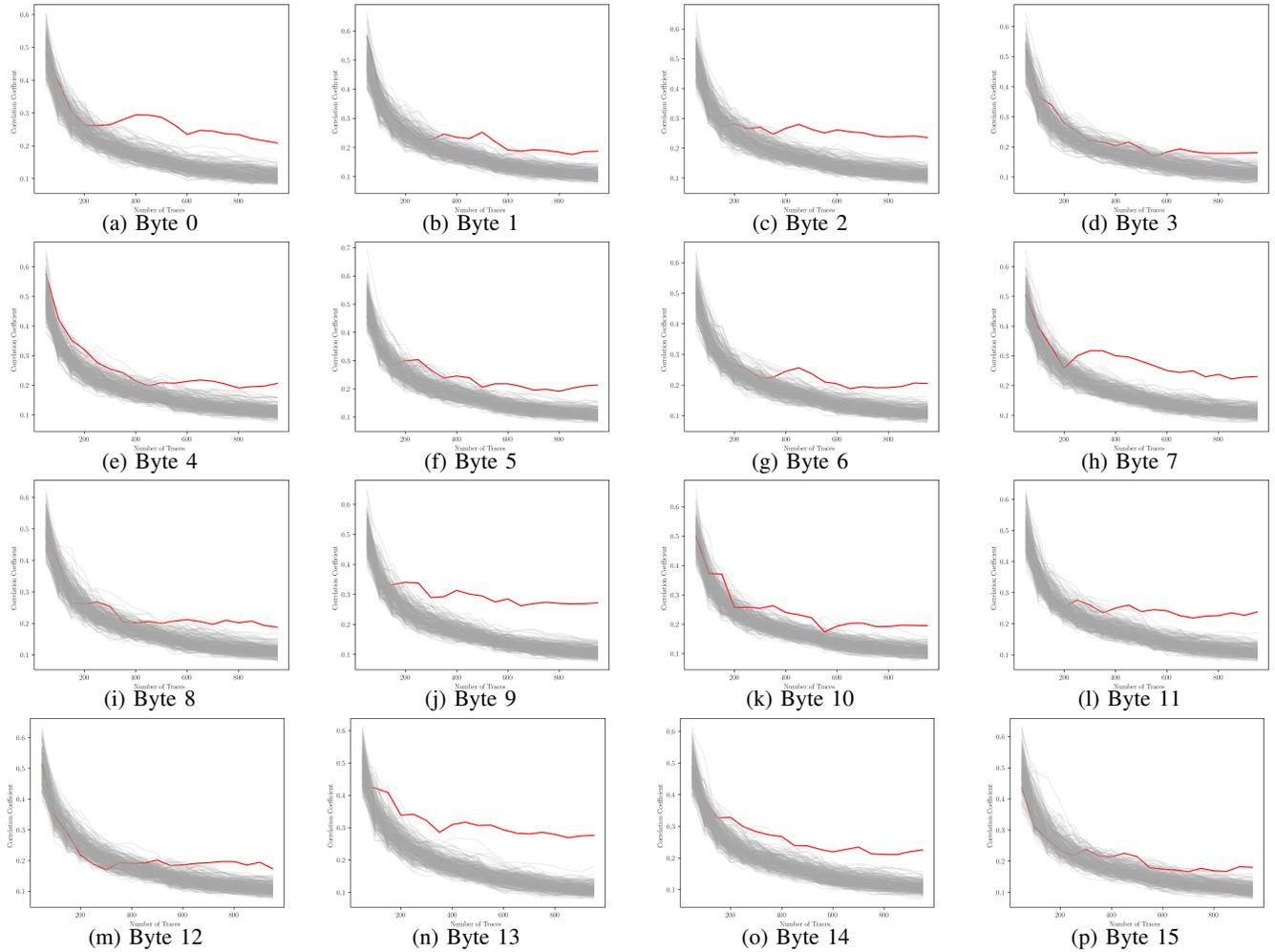}
\caption{Correlation coefficient for byte 0 to byte 15 of an AES 128-bit key with 256 different key values across varying numbers of averaged traces. Note that there is a noticeable difference between the correct key byte (indicated by the red line) and incorrect key bytes which are identified during the leakdown test.}
\label{fig:CPA_ard}
\end{figure*}

\subsubsection{Results}
Figure~\ref{fig:case_study_1_SPA} illustrates the SPA of traces captured during the experiment. Figures~\ref{fig:SPA_ard} presents the PSVC signature of AES-128 on \textit{ATmega328} MCU and Figure~\ref{fig:SPA_stm32} presents the PSVC signature of \textit{STM32F401} MCU. Since AES-128 computation consists of ten encryption rounds, the traces exhibit ten distinct power signatures corresponding to each encryption round. As illustrated in Figure~\ref{fig:SPA_stm32_1_round}, we can clearly distinguish individual functions within the AES algorithm for \textit{STM32F401}. Since the traces captured for \textit{ATmega328} exhibit more noise compared to \textit{STM32F401}, we conducted CPA for \textit{ATmega328} to simulate the worst-case scenario. The correlation coefficient results with a varying number of averaged traces are displayed in Figure~\ref{fig:CPA_ard}, which highlights two key findings. First, they provide evidence of PSVC coupling and the side-channel vulnerability of PSVC. and demonstrate the ability to mount a key recovery attack using the PSVC vulnerability. Second, they demonstrate that as the number of traces increases, the success rate of the CPA attack also increases, which is useful in the next case studies with complex scenarios and more noise.




\vspace{-0.05in}
\subsection{Case Study 2: Exploit PSVC Vulnerability to Mount an Attack using Voltage Regulator}\label{subsec:case_2}
We have created this case study to illustrate that the effect of PSVC can be reflected via secondary ICs that share the same power source. In this case study, the victim device and the attack-launching device share the same power source. For this case study, we set the adversary capability as \textit{\textbf{level-2}} as defined in the threat model (Section~\ref{subsec:Threat}). Here the adversary has physical access to the attack-launching IC but is not allowed to do any modifications to the attack-launching IC or to the victim MCU. First, we explain the experimental setup used for this case study. Then we perform an end-to-end attack and demonstrate the possibility of launching a PSVC-based side-channel attack via a neighboring IC that shares the same power source.

\subsubsection{Experimental Setup}

\begin{figure}[htp]
\centering
\vspace{-0.15in}
\input{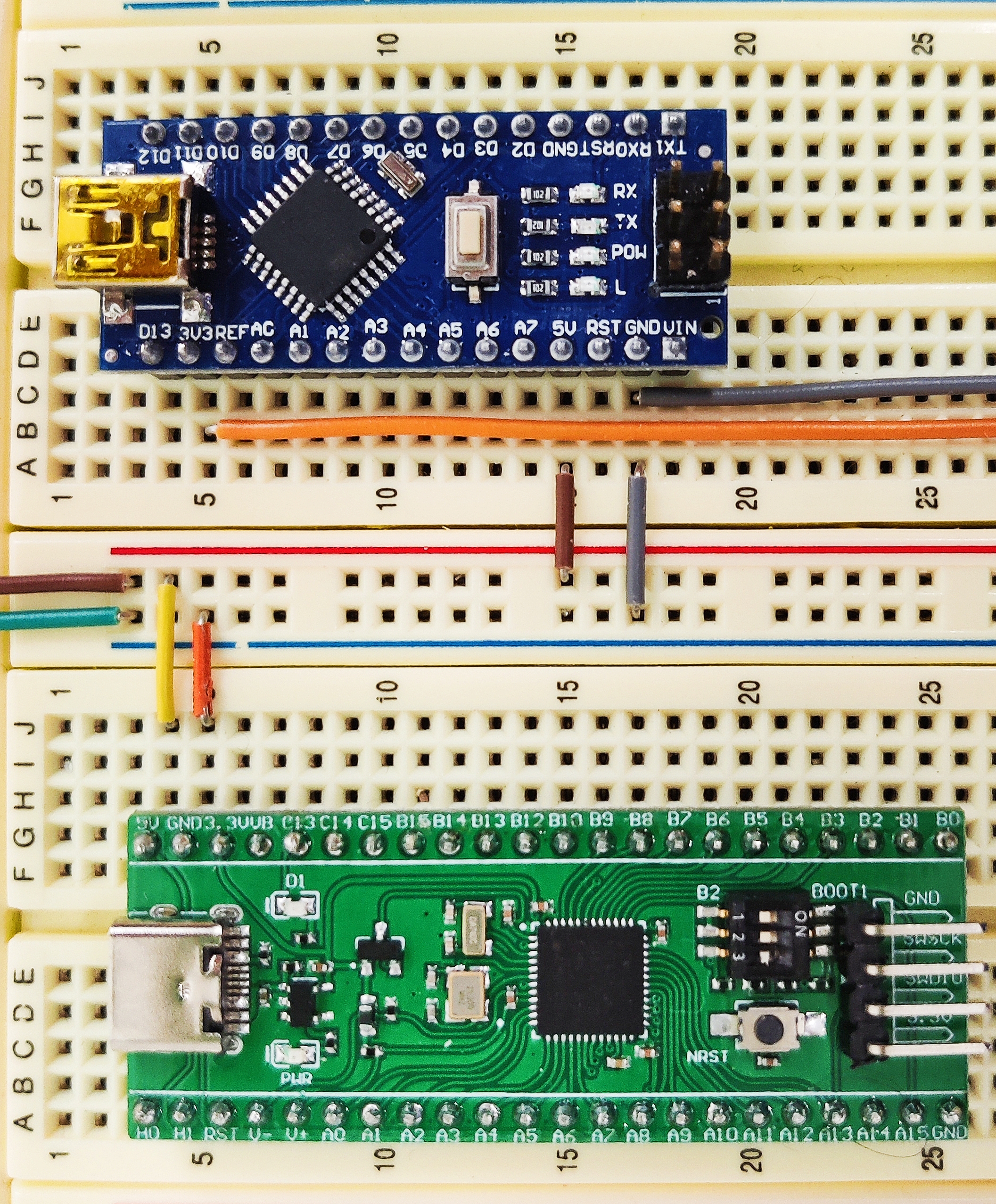}
\vspace{-0.1in}
\caption{Attack setup in Case Study 2. Both the victim IC and the attack-launching IC share the same power source. }
\label{fig:method_2}
\vspace{-0.1in}
\end{figure}

Figure~\ref{fig:method_2} shows the experimental setup for Case Study 2. Here, we have utilized the same power supply that we used in Case Study 1. For this experiment, we opted for \textit{STM32F401} MCU as the victim device, and the 3.3V Low Dropout voltage regulator of the \textit{Arduino nano} board as the attack-launching IC. We selected the Voltage Regulator Module (VRM) as the attack-launching IC in this experiment since VRMs are available in almost all devices to convert higher supply voltage into lower levels required by various electronic components~\cite{zhu2023security}. The attack methodology is similar to Case Study 1, except instead of probing victim device power rails (input voltage domain), we take measurements by probing the VRM output (regulated voltage domain).

\subsubsection{Results}

\begin{figure}[htp]
\centering
\vspace{-0.1in}
\begin{subfigure}{0.5\linewidth}
\begin{center}
\includegraphics[width=\textwidth]{fig/stm_full_enc_avg_100.jpg}
\end{center}
\vspace{-0.1in}
\caption{PSVC with direct probing}\label{fig:SPA_stm32_comp}
\end{subfigure}%
\begin{subfigure}{0.5\linewidth}
\begin{center}
\includegraphics[width=\textwidth]{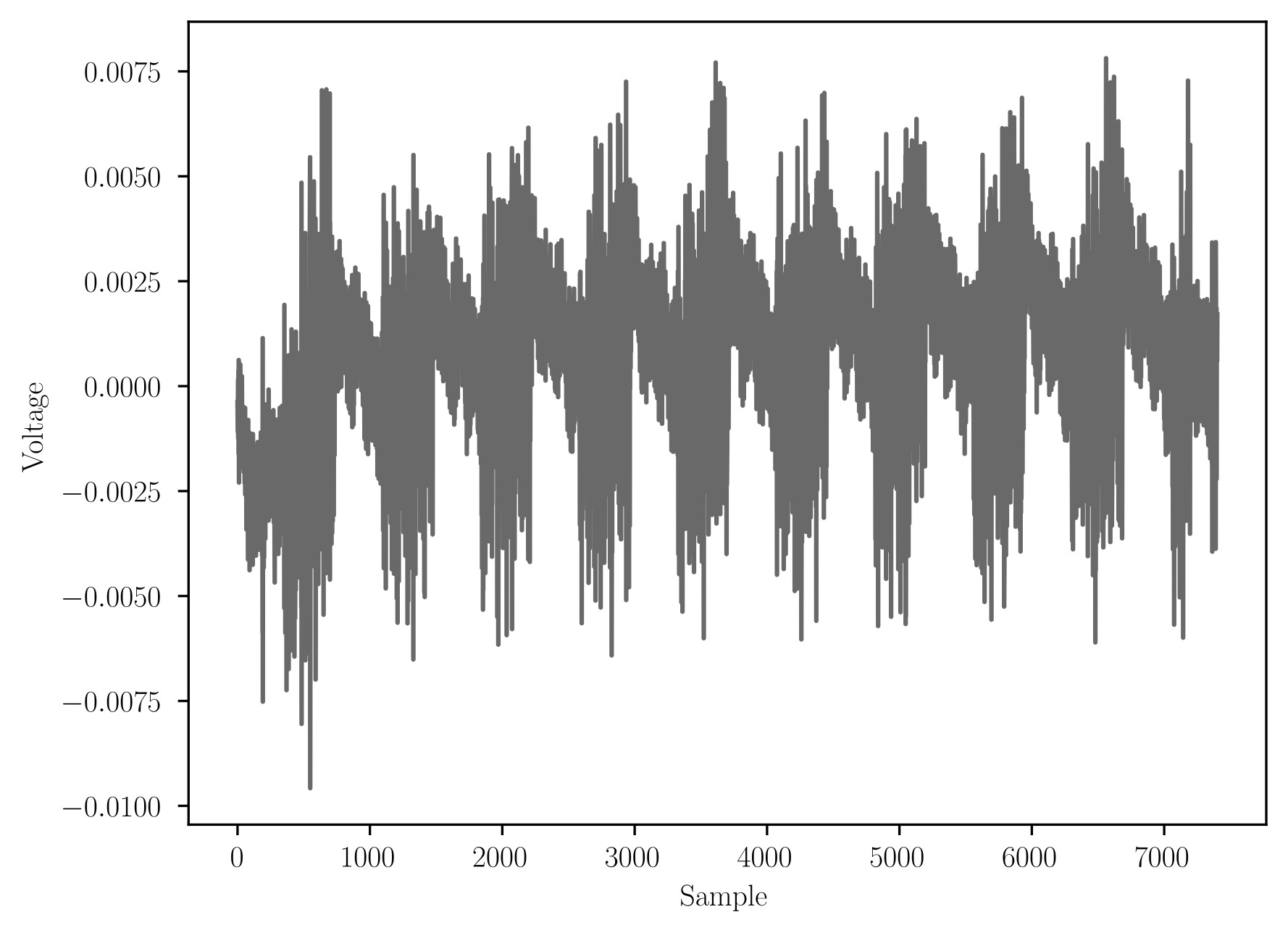}
\end{center}
\vspace{-0.1in}
\caption{PSVC with VRM probing}\label{fig:SPA_stm32_ldo}
\end{subfigure}
\vspace{-0.1in}
\caption{Comparison between PSVC traces of \textit{STM32F401} collected by direct probing into the power rails vs PSVC trace collected from the $3.3V$ VRM output of the \textit{Arduino nano}. Although in Figure~\ref{fig:SPA_stm32_ldo}, the trace is collected in another voltage domain of the same circuit, the PSVC signature is still preserved compared to the original domain (Figure~\ref{fig:SPA_stm32_comp}).}
\label{fig:ldo_out_trace_spa}
\vspace{-0.1in}
\end{figure}

Figure~\ref{fig:ldo_out_trace_spa} illustrates the comparison between the SPA of captured traces for AES-128 on the \textit{STM32F401} through direct probing (Figure~\ref{fig:SPA_stm32_comp}) of the power rails versus traces captured through the $3.3V$ output of the \textit{Arduino nano} via the VRM (Figure~\ref{fig:SPA_stm32_ldo}). 

Similar to Figure~\ref{fig:SPA_stm32_comp}, ten rounds of AES-128 implementation are clearly visible in Figure~\ref{fig:SPA_stm32_ldo}. However, the noise level of the captured traces is elevated due to interference from various components. Our findings in this case study reveal that PSVC can propagate through VRMs from one voltage domain to another within an electronic device.
Although the major responsibility of VRMs is isolating voltage domains, effectively preventing the propagation of voltage fluctuations between them, PSVC side-channel vulnerability exploits the tolerable noise margin accepted by the VRM and propagates the PSVC signature across voltage domains.


\subsection{Case Study 3: Exploit PSVC Vulnerability to Mount an End-to-End Wireless Attack}\label{subsec:case_3}

In this section, we perform a completely remote end-to-end attack exploiting the PSVC vulnerability on the victim device which consists of a wireless Bluetooth module. For this case study, we set the adversary capability as \textit{\textbf{level-3}} as defined in the threat model (Section~\ref{subsec:Threat}). Here the adversary is within the wireless range of the victim's device. First, we outline the experimental setup. Next, we perform an end-to-end remote attack and demonstrate the attack capabilities of PSVC-based side-channel attack when the adversary does not have physical access to the device under attack.

\begin{figure}[htp]
\centering
\vspace{-0.1in}
\include{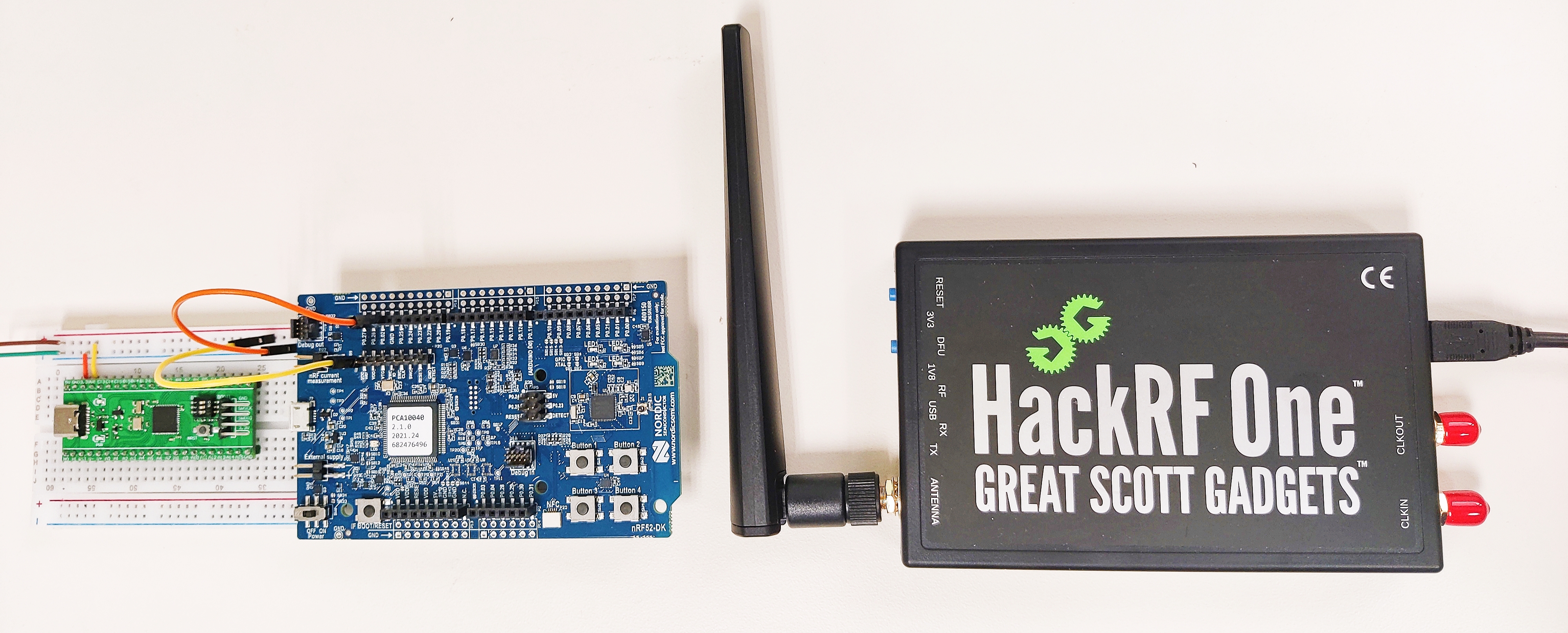}
\vspace{-0.3in}
\caption{Attack setup used in the Case Study 3. Here the PSVC traces are collected from the HackRF~\cite{HackRF} SDR.}
\label{fig:method_3}
\end{figure}

\subsubsection{Experimental Setup}

Figure~\ref{fig:method_3} presents the experimental setup we have used for Case Study 3. Here the device under attack consists of two physically separated development boards that share the same power supply. We have the \textit{BlackPill} development board which runs the cryptographic application (APP) and the \textit{NRF52-DK} development board consists of a programmable \textit{nRF52832} Bluetooth transceiver that runs a generic Bluetooth transmission. In this experiment, we utilized the radio test example, which provides support for nRF52-DK, and made modifications to enable the transmission of a high-power continuous carrier signal at 2.4 GHz. Both \textit{BlackPill} and the \textit{nRF52-DK} boards were powered using the same power supply. During the experiment, we maintained a distance of approximately 50 cm between the two boards to minimize electromagnetic coupling. As a result, the sole connection between the \textit{BlackPill} and \textit{nRF52-DK} was through the shared power lines. Instead of probing the power rails with the oscilloscope, in this experiment, we use the HackRF~\cite{HackRF} software-defined radio module to capture the Bluetooth transmission of the \textit{NRF52-DK} development board via CubicSDR application. We placed the HackRF at a small distance (10 cm) from the \textit{NRF52-DK}. After capturing the RF signal, we convert the frequency domain into the time domain and apply the steps discussed in Section~\ref{sec:methodology}.

\subsubsection{Results}

\begin{figure}[htp]
\centering
\vspace{-0.05in}
\includegraphics[width=0.9\linewidth]{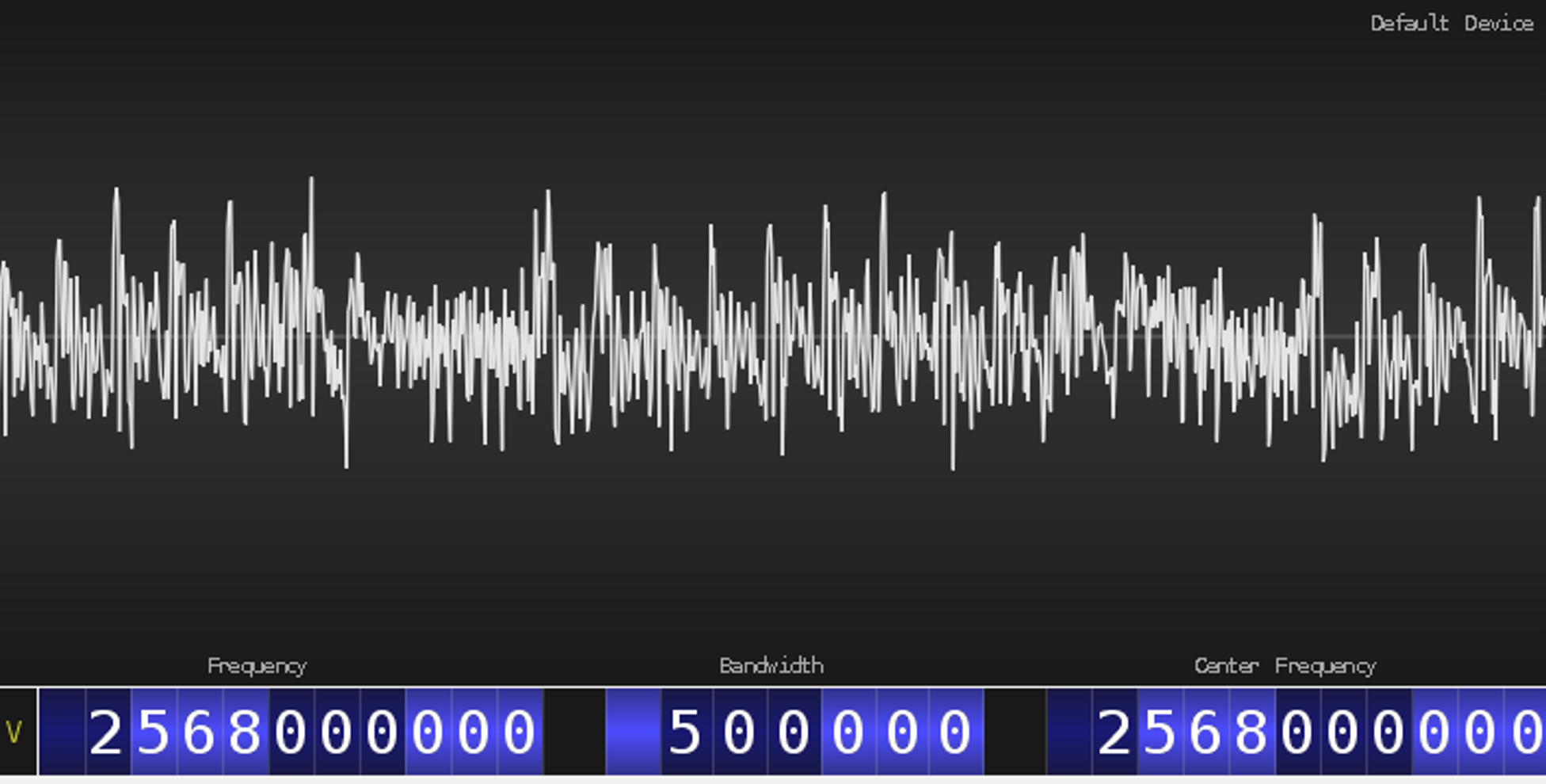}
\caption{ The signature of PSVC leakage of the \textit{BlackPill} development board observed through  CubicSDR~\cite{CubicSDR} application. The trace capture was performed with a HackRF~\cite{HackRF} SDR  tuned to a harmonic of the \textit{STM32F401} clock frequency which receives the amplitude-modulated (AM) Bluetooth carrier signal. Here, we can identify the rounds of AES-128. Note that it may not be as clear as the SPA of previous case studies.}
\label{fig:PSVC_RF}
\end{figure}

Figure~\ref{fig:PSVC_RF} displays the SPA of the captured trace obtained using HackRF and CubicSDR. In comparison to the SPA of previous case studies, this SPA exhibits more noise. Nonetheless, we can still recover the signature of ten rounds of AES-128 within the trace. 

\begin{figure}[htp]
\centering
\vspace{-0.1in}
\begin{subfigure}{0.5\linewidth}
\begin{center}
\includegraphics[width=\textwidth]{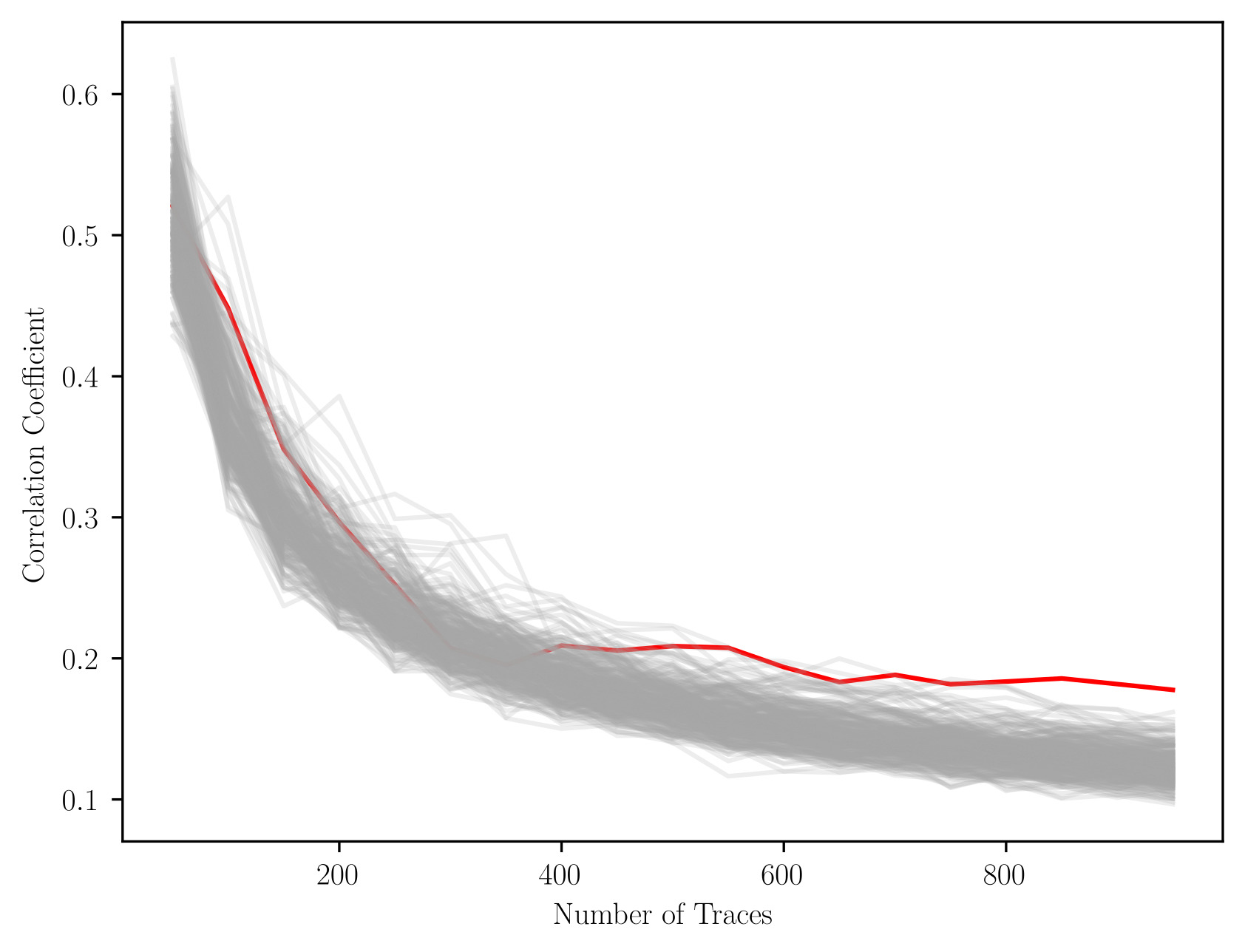}
\end{center}
\vspace{-0.1in}
\caption{Byte 0}\label{fig:wireless_1}
\end{subfigure}%
\begin{subfigure}{0.5\linewidth}
\begin{center}
\includegraphics[width=\textwidth]{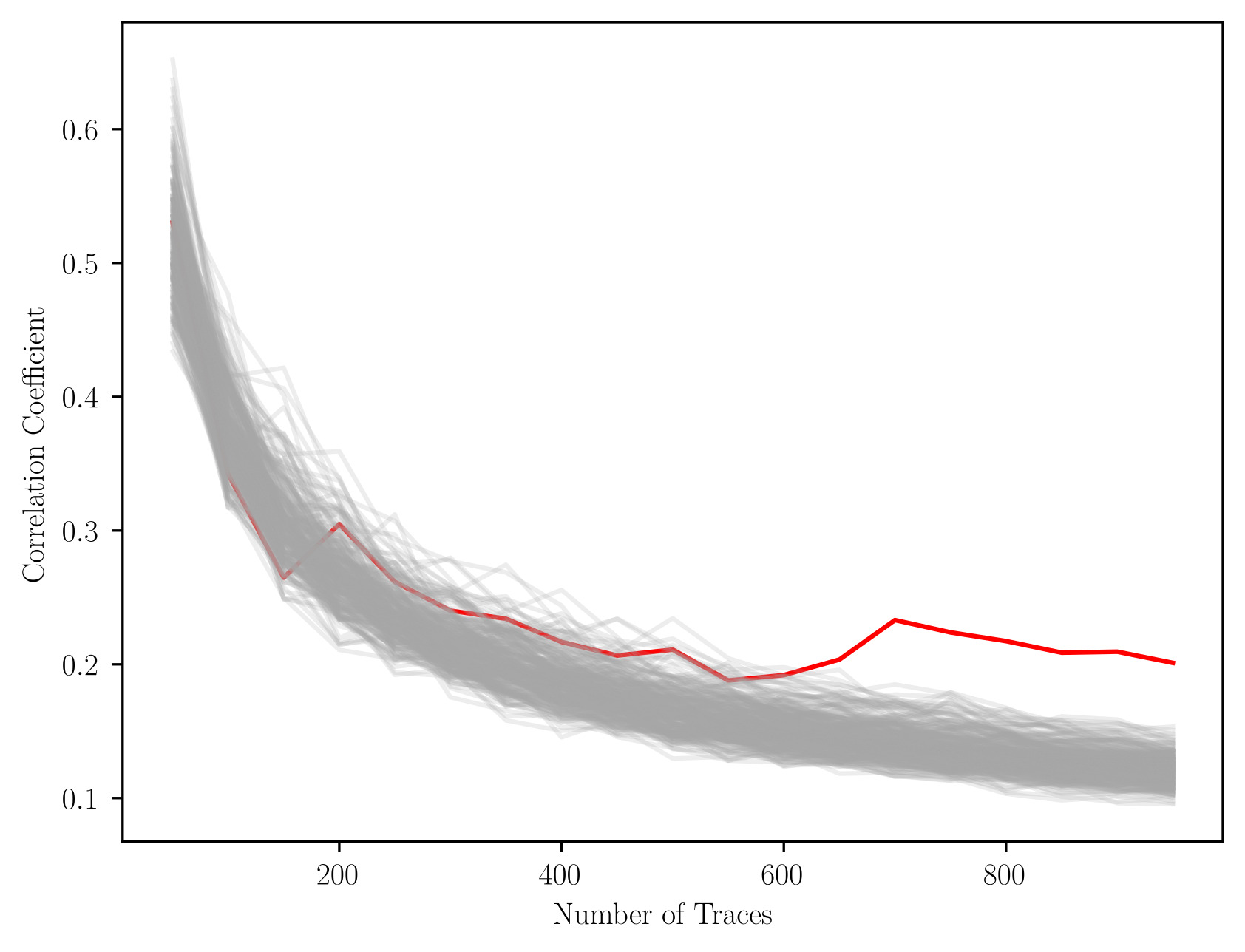}
\end{center}
\vspace{-0.1in}
\caption{Byte 15}\label{fig:wireless_2}
\end{subfigure}
\vspace{-0.2in}
\caption{Results of the key bytes recovery after performing the leakdown test on the end-to-end attack launched on \textit{BlackPill} development board via the Bluetooth carrier signal of the \textit{nRF52 DK} development board. Traces are captured using the HackRF SDR module and recorded by the CubicSDR application followed by filtering with $Avg(300)$.}
\label{fig:case-study-3}
\end{figure}

Figure~\ref{fig:case-study-3} presents leakdown test results after performing CPA on two key bytes (Byte 0 in Figure~\ref{fig:wireless_1} and Byte 15 in Figure~\ref{fig:wireless_2}) of end-to-end remote attack on \textit{BlackPill} development board via a Bluetooth carrier signal. 

\begin{figure*}[t]
\begin{subfigure}{0.32\linewidth}
\begin{center}
\includegraphics[width=\textwidth]{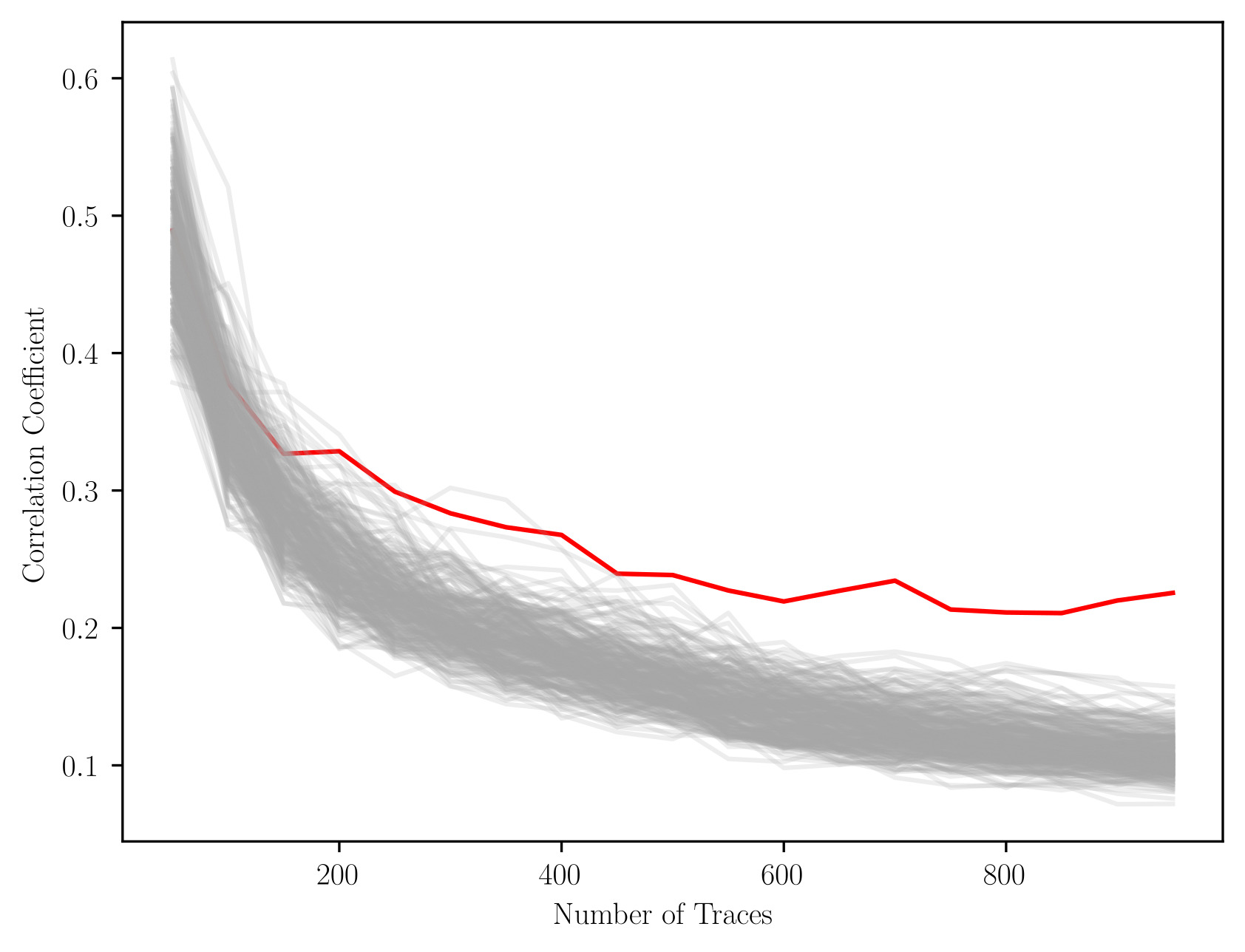}
\end{center}
\vspace{-0.1in}
\caption{Input voltage = $5.0V$}\label{fig:cs4-5}
\end{subfigure}%
\hspace{0.1in}
\begin{subfigure}{0.32\linewidth}
\begin{center}
\includegraphics[width=\textwidth]{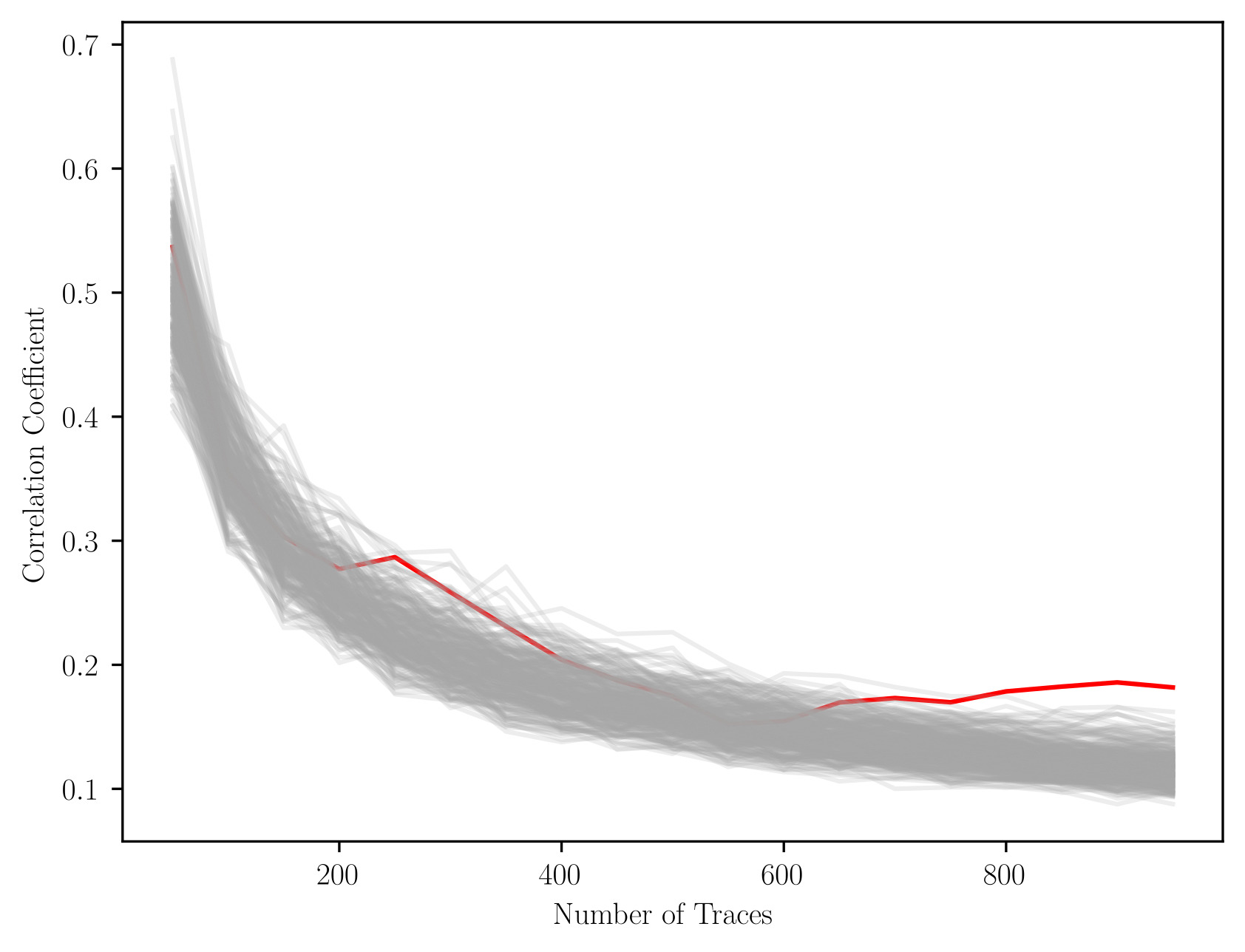}
\end{center}
\vspace{-0.1in}
\caption{Input voltage = $4.0V$}\label{fig:cs4-4}
\end{subfigure}
\hspace{0.1in}
\begin{subfigure}{0.32\linewidth}
\begin{center}
\includegraphics[width=\textwidth]{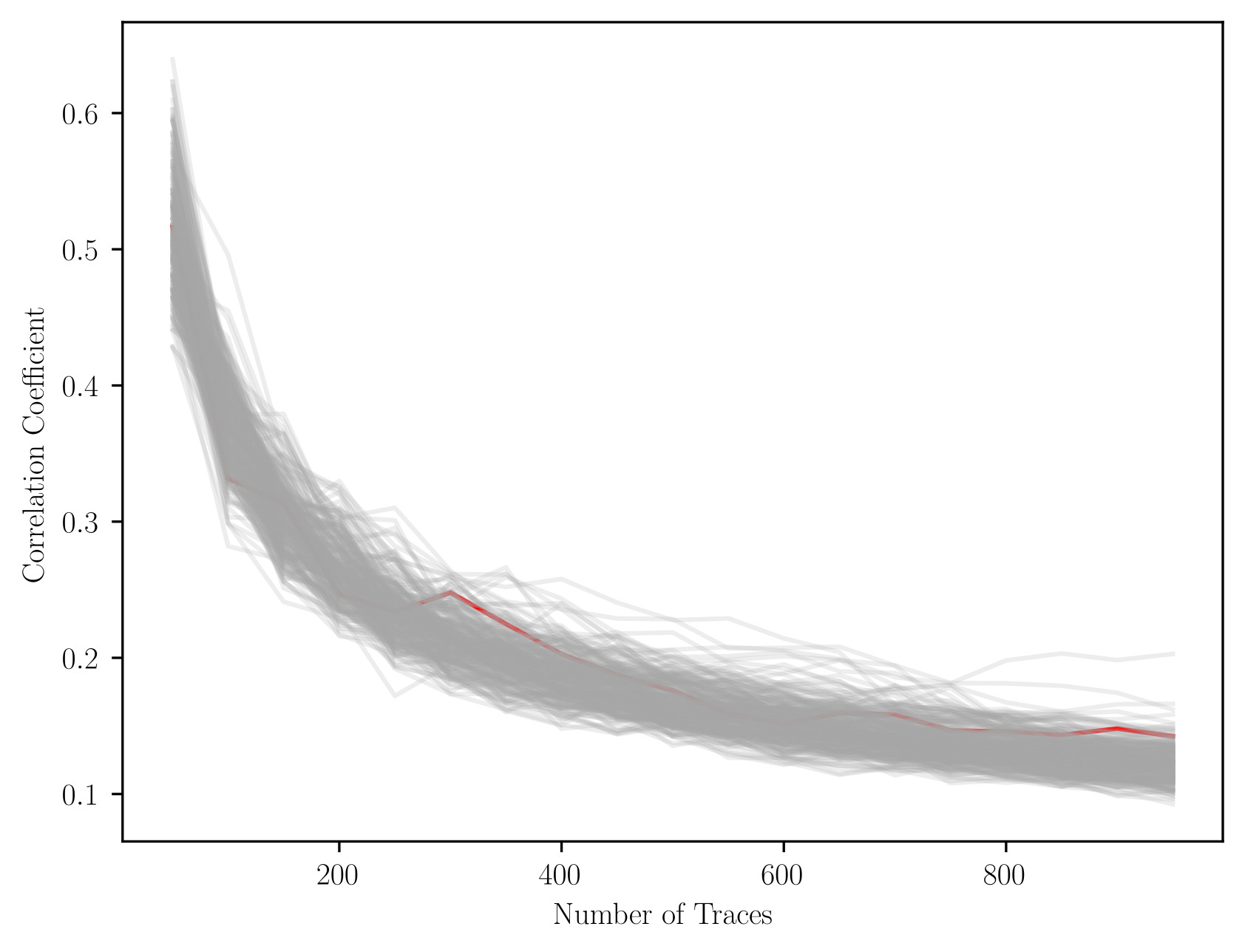}
\end{center}
\vspace{-0.1in}
\caption{Input voltage = $3.0V$}\label{fig:cs4-3}
\end{subfigure}
\vspace{-0.2in}
\caption{Correlation coefficient for example key byte of an AES 128-bit key with 256 different key values across varying numbers of averaged traces for different input voltages. The correct key byte is indicated by the red line.}
\vspace{-0.15in}
\label{fig:corr_with_diff_vcc}
\end{figure*}

This leads to the conclusion that if the PSVC signature is strong enough, it can propagate from the victim MCU to the attack IC (in this case, RF). Therefore, the results of our experiments from Case Study 1, Case Study 2, and Case Study 3 provide evidence of the following:
(1) the existence of PSVC, (2) the potential of PSVC as a vulnerability, (3) the ability to propagate from one voltage domain to another using the same power source, and (4) the ability to propagate from the digital domain (MCU) to the analog domain (RF).

\vspace{-0.05in}
\subsection{Case Study 4: Effect of Supply Voltage Range on PSVC Attack}
\label{subsec:case_4}

In this section, we explore the effect of the supply voltage range of the power source on the recoverability of secret data from the application (APP) with a PSVC side-channel attack. For this, we change the quality of the power supply with different voltage levels supported by the device under attack. In this section, we first outline the experimental setup. Next, we present the results about the behavior of PSVC side-channel leakage with the variation of input voltage.

\subsubsection{Experimental Setup}

We use the \textit{Arduino nano} development board with the \textit{Atmega328} MCU and power it with a variable power supply. \textit{Atmega328} MCU has a rated input voltage range between $+1.8V$ to $+5.5V$. For this case study, we use the same setup of Case Study 1 for different input voltages of $3.0V, 4.0V$, and $5.0V$.


\subsubsection{Results}

\begin{figure}[htp]
\centering
\vspace{-0.1in}
\includegraphics[width=0.48\textwidth]{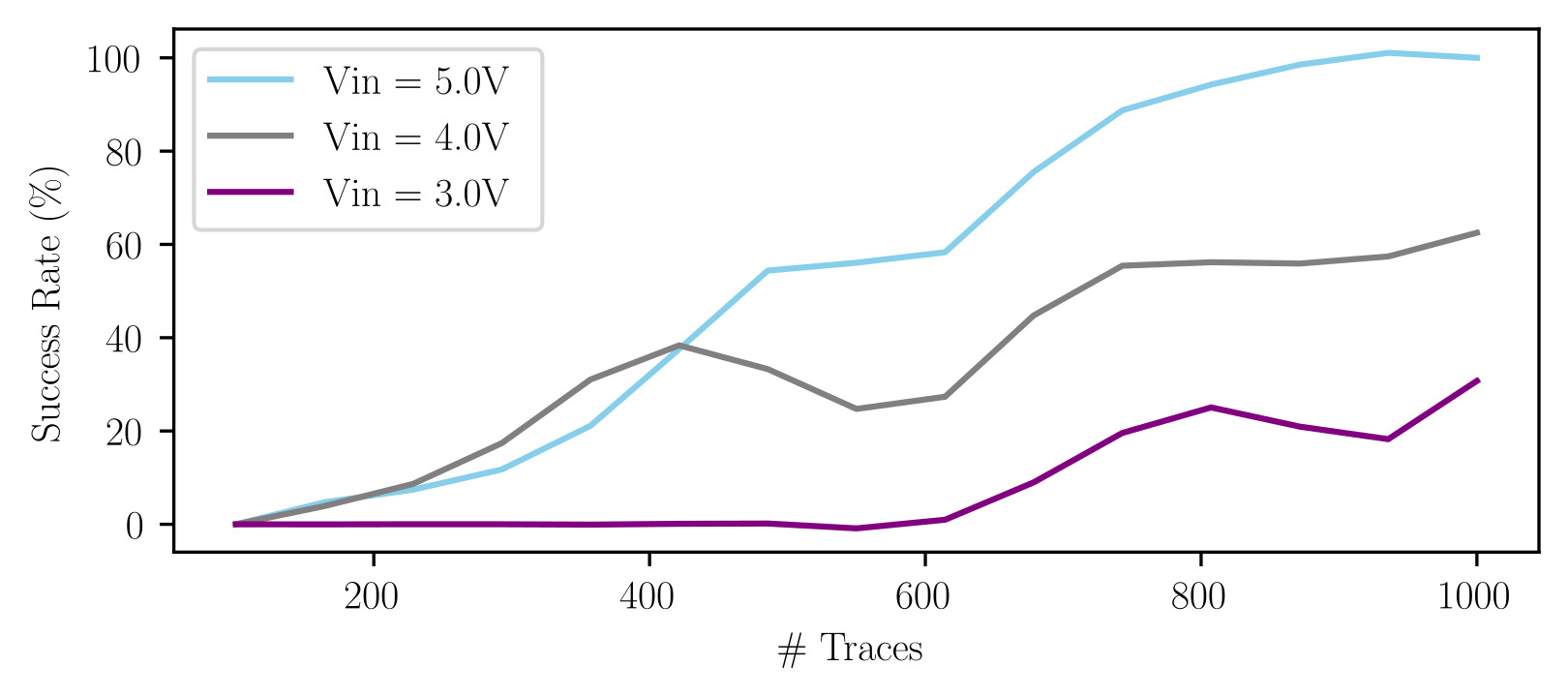}
\vspace{-0.1in}
\caption{SCA success rate with different input voltages. It can be observed when the device is operating at a higher input voltage ($5V$), the attack success rate is higher and the number of traces to perform an attack is less compared to the same device operating at a lower input voltage ($3V$).}
\label{fig:success_rate_vs_v_in}
\end{figure}

Figure~\ref{fig:corr_with_diff_vcc} illustrates the leakdown test results after performing CPA for input voltages of $5.0V, 4.0V$ and $3.0V$ in Figure~\ref{fig:cs4-5}, Figure~\ref{fig:cs4-4} and Figure~\ref{fig:cs4-3}, respectively. Figure~\ref{fig:success_rate_vs_v_in} presents the attack success rate after performing the experiment for three selected input voltage values. The behavior illustrated in Figure~\ref{fig:success_rate_vs_v_in} can be primarily attributed to variations in the signal-to-noise ratio (SNR) with the input supply voltage. 
Our observations revealed a decline in the SNR of the captured trace as the input voltage decreases, as demonstrated in Table~\ref{tab:SNR_vs_Vin}. This reduction in SNR can be mainly attributed to the reduced strength of power fluctuations at lower input voltages. It is important to highlight that all input voltage values used in our experiments remained within the specified operating voltage range of the \textit{ATmega 328} MCU. 

This experiment highlights two significant facts. First, our results demonstrate that operating an MCU at its minimum input voltage is a preferable choice for enhancing security, as it is more challenging to attack the MCU when the input voltage is near the lower bound of the specified operating voltage. Second, in the context of battery-operated electronic devices, similar voltage reductions may be observed over time due to battery aging. Consequently, the input voltage supplied by the battery gradually decreases, mirroring the numerical trends in our experiment. Therefore, we can conclude that as electronic devices age, they become more resilient against PSVC-based attacks.

\begin{table}[htp]
    \centering
        \caption{SNR of captured traces with different input voltages (compared to the power trace captured using the conventional method, as shown in Figure~\ref{fig:existing})}
    \begin{tabular}{|l|c|c|c|}
\hline
Vin (V)  & 3   & 4   & 5  \\ \hline
SNR (dB) & \cellcolor[HTML]{FFE599}-9.77 & \cellcolor[HTML]{D9EAD3}-6.53 & \cellcolor[HTML]{38761D}{\color[HTML]{FFFFFF} -5.95} \\ \hline
\end{tabular}

    \label{tab:SNR_vs_Vin}
    \vspace{-0.1in}
\end{table}

\section{Conclusion}\label{sec:conclusion}

In this paper, we introduced a novel vulnerability that can leak sensitive information from physical layer supply voltage coupling (PSVC) on off-the-shelf devices. We discussed the reasons behind the PSVC side-channel leakage and proposed a methodology to evaluate different device configurations against PSVC vulnerability. We conducted four different case study experiments, where the first three case studies were focused on launching end-to-end attacks exploiting the PSVC vulnerability with different adversary capabilities and different device configurations. Experimental results revealed that PSVC vulnerability is present in off-the-shelf hardware components, and can be exploited to mount attacks in multiple ways including an extreme remote attack where the adversary does not need physical access to the device under attack. In the final case study, we performed an end-to-end attack while changing the supply voltage in the manufacturer-specified range.
This experiment revealed that the effect of PSVC vulnerability in different voltage levels is affected by the signal-to-noise ratio and when the device is operating at the lowest specified operational voltage, performing an attack using the PSVC vulnerability requires much more effort than performing an attack in the highest voltage level. The same effect of reduced risk of PSVC vulnerability should be observed on battery-operated devices, with the battery aging over time. We also discussed potential methods that designers can follow to reduce the effect of PSVC leakage. In conclusion, PSVC leakage can hide within the manufacturing tolerances of the components and therefore requires considerable attention when designing secure and trustworthy devices.





\section*{Acknowledgments}
This work was partially supported by the grant from the National Science Foundation (CCF-1908131).


\bibliographystyle{IEEEtran}
\bibliography{IEEEabrv,bibliography}

\vspace{-0.5in}
\begin{IEEEbiography}
[{\includegraphics[width=1in,clip]{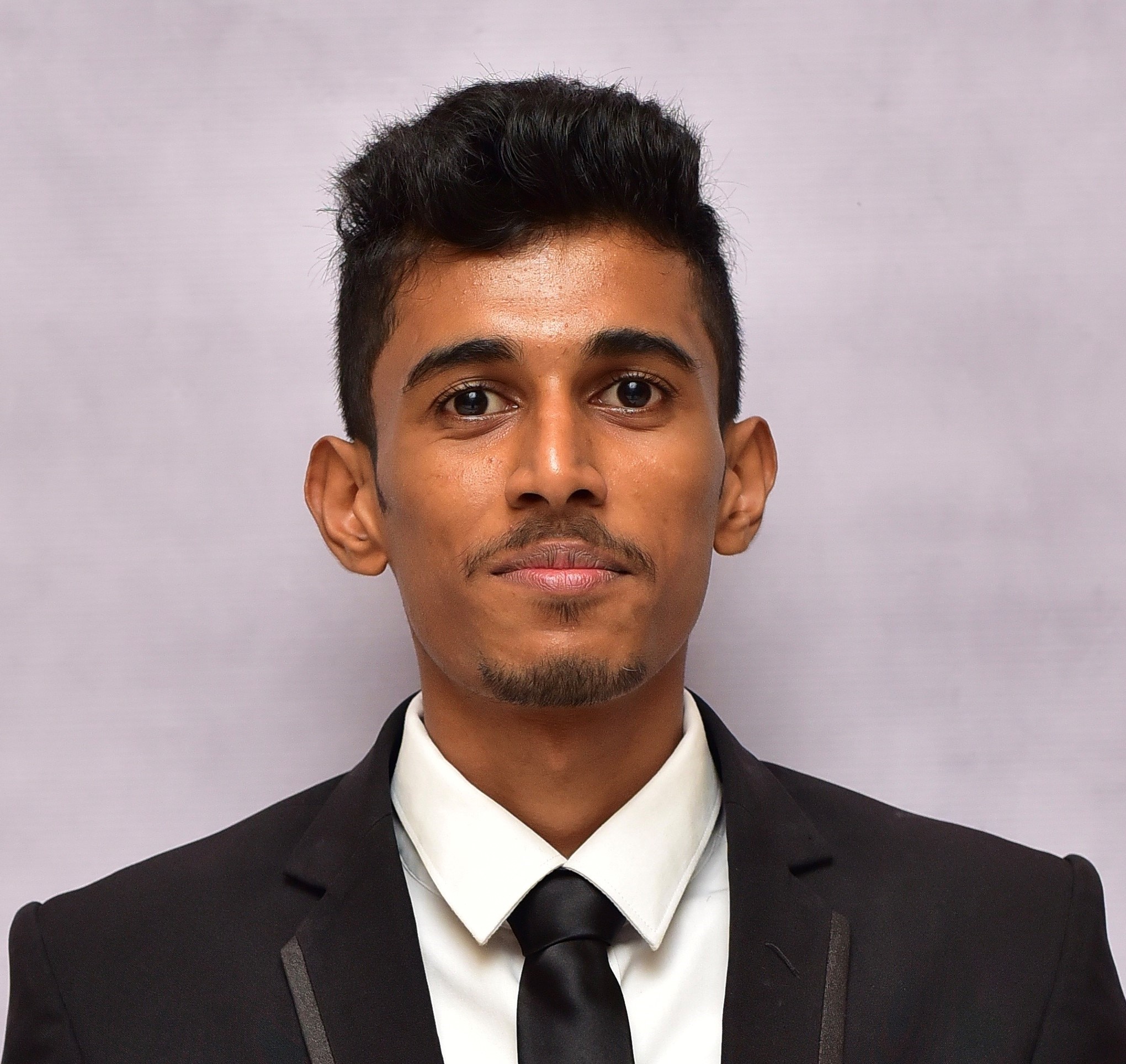}}]{Sahan Sanjaya} is a second-year Ph.D student in the Department of Computer \& Information Science \& Engineering at the University of Florida. In 2022, he completed his B.Sc. in the Department of Electronic and Telecommunication Engineering at the University of Moratuwa, Sri Lanka. His research interests encompass side-channel attacks, hardware security, pre-silicon validation, and post-silicon validation.
\end{IEEEbiography}

\vspace{-0.5in}
\begin{IEEEbiography}[{\includegraphics[width=1in,clip]{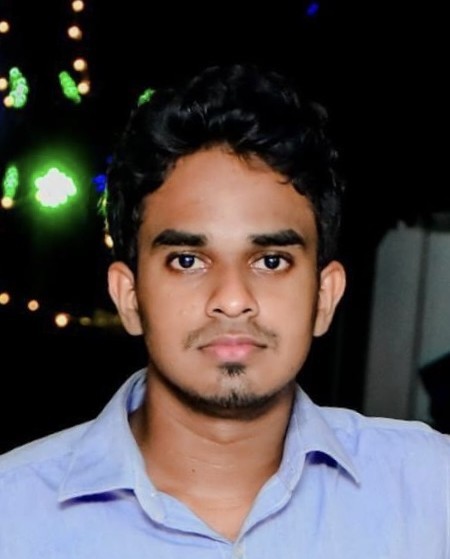}}]{Aruna Jayasena} is a Ph.D student in the Department of Computer \& Information Science \& Engineering at the University of Florida. He received his B.S. in the Department of Computer Science and Engineering at the University of Moratuwa, Sri Lanka, in 2019. His research focuses on systems security, hardware-firmware co-validation, test generation, trusted execution, side-channel analysis, and system-on-chip debug.
\end{IEEEbiography}

\vspace{-6.5in}
\begin{IEEEbiography}[{\includegraphics[width=1in,clip,keepaspectratio]{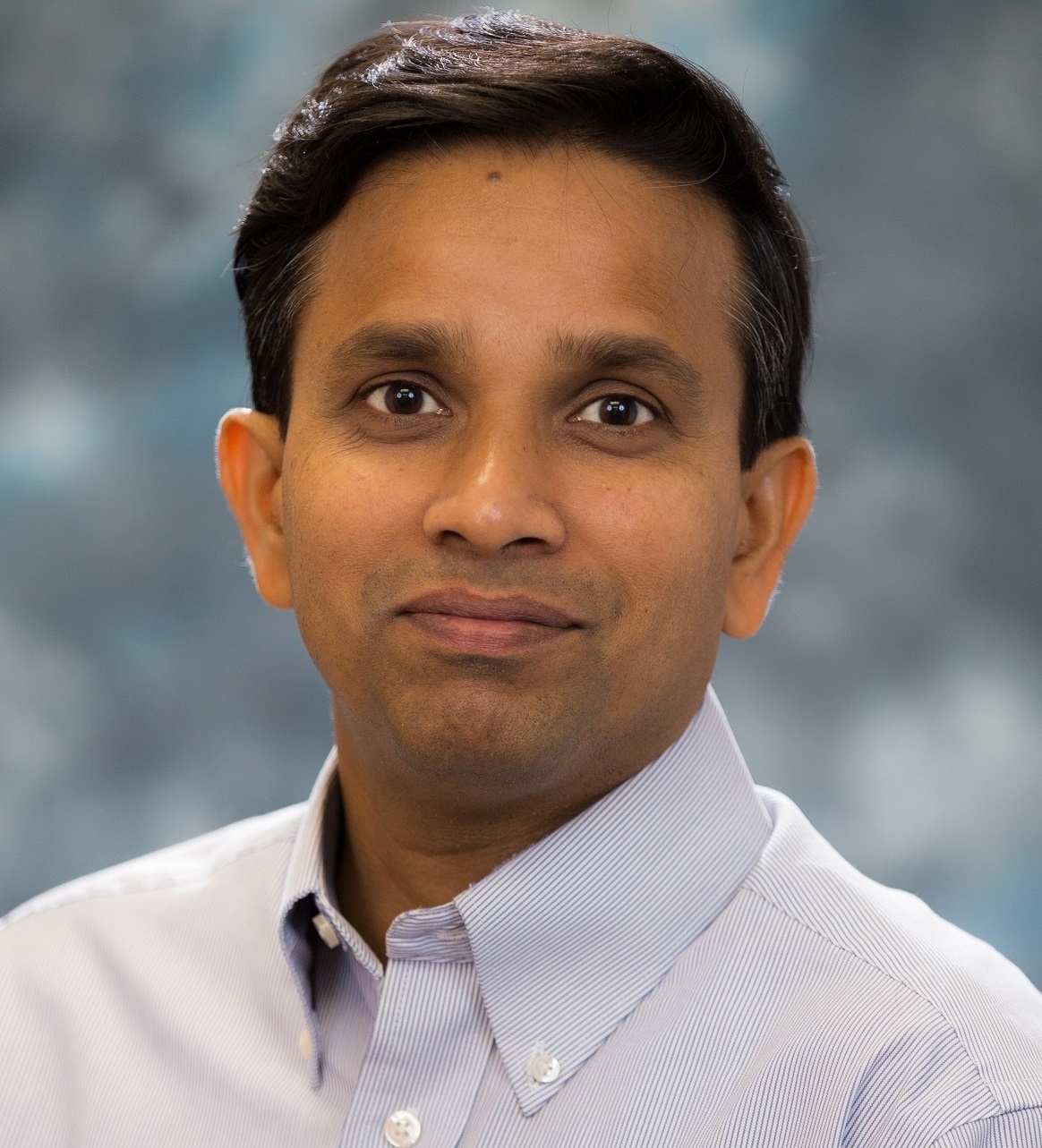}}]{Prabhat Mishra}
is a Professor in the Department of Computer and Information Science and Engineering at the University of Florida. He received his Ph.D. in Computer Science from the University of California at Irvine in 2004. His research interests include embedded and cyber-physical systems, hardware security and trust, and energy-aware computing. He currently serves as an Associate Editor of IEEE Transactions on VLSI Systems and ACM Transactions on Embedded Computing Systems. He is an IEEE Fellow and an ACM Distinguished Scientist.
\end{IEEEbiography}

\end{document}